\documentclass[sigconf, nonacm]{acmart}
\usepackage[utf8]{inputenc}
\usepackage[utf8]{inputenc}
\DeclareUnicodeCharacter{202F}{\,}

\usepackage{algorithm}
\usepackage{algpseudocode}
\usepackage{subcaption}
\usepackage{placeins}
\usepackage{diagbox}
\usepackage{array}
\usepackage[table]{xcolor}  % For coloring
\usepackage{booktabs}
\usepackage{caption}
\usepackage[utf8]{inputenc}
\usepackage{tikz}
\usepackage[symbol]{footmisc}
\usepackage{float}
\usepackage{needspace}
\usepackage{enumitem}
\usetikzlibrary{shapes.geometric, arrows.meta, positioning, calc, patterns}

\newcommand\vldbdoi{XX.XX/XXX.XX}
\newcommand\vldbpages{XXX-XXX}
\newcommand\vldbvolume{14}
\newcommand\vldbissue{1}
\newcommand\vldbyear{2020}
\newcommand\vldbauthors{\authors}
\newcommand\vldbtitle{\shorttitle} 
\newcommand\vldbavailabilityurl{URL_TO_YOUR_ARTIFACTS}
\newcommand\vldbpagestyle{plain} 

\begin{document}
\title{QVCache: A Query-Aware Vector Cache}

%%
%% The "author" command and its associated commands are used to define the authors and their affiliations.
\author{Anıl Eren Göçer}
\affiliation{%
  \institution{ETH Zurich}
  \city{Zurich}
  \country{Switzerland}
}
\email{agoecer@ethz.ch}

\author{Ioanna Tsakalidou}
\affiliation{%
  \institution{EPFL}
  \city{Lausanne}
  \state{Switzerland}
}
\email{ioanna.tsakalidou@epfl.ch}

\author{Hamish Nicholson}
\affiliation{%
  \institution{EPFL}
  \city{Lausanne}
  \state{Switzerland}
}
\email{hamish.nicholson@epfl.ch}

\author{Kyoungmin Kim}
\affiliation{%
  \institution{EPFL}
  \city{Lausanne}
  \state{Switzerland}
}
\email{kyoung-min.kim@epfl.ch}

\author{Anastasia Ailamaki}
\affiliation{%
  \institution{EPFL}
  \city{Lausanne}
  \state{Switzerland}
}
\email{anastasia.ailamaki@epfl.ch}


\begin{abstract}
Vector databases have become a cornerstone of modern information retrieval, powering applications in recommendation, search, and retrieval-augmented generation (RAG) pipelines. However, scaling Approximate Nearest Neighbor (ANN) search to high recall under strict latency SLOs remains fundamentally constrained by memory capacity and I/O bandwidth. Disk-based vector search systems suffer severe latency degradation at high accuracy, while fully in-memory solutions incur prohibitive memory costs at billion-scale. Despite the central role of caching in traditional databases, vector search lacks a general query-level caching layer capable of amortizing repeated query work.

We present QVCache, the first backend-agnostic, query-level caching system for ANN search with bounded memory footprint. QVCache exploits semantic query repetition by performing similarity-aware caching rather than exact-match lookup. It dynamically learns region-specific distance thresholds using an online learning algorithm, enabling recall-preserving cache hits while bounding lookup latency and memory usage independently of dataset size. QVCache operates as a drop-in layer for existing vector databases. It maintains a megabyte-scale memory footprint and achieves sub-millisecond cache-hit latency, reducing end-to-end query latency by up to 40–1000$\times$ when integrated with existing ANN systems. For workloads exhibiting temporal-semantic locality, QVCache substantially reduces latency while preserving recall comparable to the underlying ANN backend, establishing it as a missing but essential caching layer for scalable vector search.
\end{abstract}

\maketitle

\pagestyle{\vldbpagestyle}
\begingroup\small\noindent\raggedright\textbf{PVLDB Reference Format:}\\
\vldbauthors. \vldbtitle. PVLDB, \vldbvolume(\vldbissue): \vldbpages, \vldbyear.\\
\href{https://doi.org/\vldbdoi}{doi:\vldbdoi}
\endgroup
\begingroup
\renewcommand\thefootnote{}\footnote{\noindent
This work is licensed under the Creative Commons BY-NC-ND 4.0 International License. Visit \url{https://creativecommons.org/licenses/by-nc-nd/4.0/} to view a copy of this license. For any use beyond those covered by this license, obtain permission by emailing \href{mailto:info@vldb.org}{info@vldb.org}. Copyright is held by the owner/author(s). Publication rights licensed to the VLDB Endowment. \\
\raggedright Proceedings of the VLDB Endowment, Vol. \vldbvolume, No. \vldbissue\ %
ISSN 2150-8097. \\
\href{https://doi.org/\vldbdoi}{doi:\vldbdoi} \\
}\addtocounter{footnote}{-1}\endgroup

\ifdefempty{\vldbavailabilityurl}{}{
\vspace{.3cm}
\begingroup\small\noindent\raggedright\textbf{PVLDB Artifact Availability:}\\
The source code, data, and/or other artifacts have been made available at \url{\vldbavailabilityurl}.
\endgroup
}

\section{Introduction}
%Needs to answer:
%1)necessity, whats the problem
%2)impact of the solution
%3)applicability to today's issues

% introduce the topic
Modern data-intensive services increasingly rely on Approximate Nearest Neighbor (ANN) search over high-dimensional embeddings. However, scaling ANN search to high recall under strict latency SLOs remains fundamentally constrained by memory capacity and I/O bandwidth.
This tension is now a dominant systems bottleneck: improving recall predictably inflates latency and operational cost, while controlling latency requires either aggressive approximation or prohibitively large memory footprints. As a result, ANN has become a first-order performance and cost concern in production systems, including recommendation pipelines~\cite{Covington:2016:DNN:YouTube}, search engines~\cite{Liu:2021:Q2S:Facebook, amazon-shop-the-look, visrel}, and Retrieval-Augmented Generation (RAG) workloads for large language models (LLMs)~\cite{Lewis:2020:RAG:NeurIPS, gao2024retrievalaugmentedgenerationlargelanguage}. 

% the problem
This bottleneck is structural rather than incidental. Achieving high recall in ANN search requires traversing increasingly large and irregular neighborhoods in high-dimensional space, leading to random access amplification that cannot be efficiently prefetched or batched. In graph-based indexes, recall improvements translate directly into expanded graph exploration, candidate explosion, and deeper disk or memory accesses. 
%Importantly, this tradeoff is not specific to a particular index structure: graph-based, quantization-based, and tree-based ANN indexes all incur increasing random access as recall increases. 
Although in-memory ANN systems achieve low latency, their memory cost grows linearly with the size of the dataset. Disk-based systems reduce memory pressure, but suffer severe latency degradation at high recall and large result sets (k), especially on very large datasets such as billion-scale collections, where pointer chasing and random I/O \cite{jaiswal2022ooddiskannefficientscalablegraph} dominate execution time.  As a result, operators are forced into an undesirable choice between cost and performance, with no mechanism to amortize repeated query work over time.

% enabler
A natural response to repeated expensive computation is caching. Across the systems stack, caching has been the primary mechanism for amortizing work. %from hardware memory hierarchies~\cite{memory-hierarchy} to database buffer pools~\cite{dbmin,data-caching-for-enterprise-scale}, intermediate results caching~\cite{caching-intermediate-results}, and final query result caches~\cite{memcached,redis2025}. 
However, existing techniques rely on a core assumption: identical queries recur. This assumption underlies the cache abstraction itself, which treats queries as discrete keys rather than points in a metric space. However, in vector search, this assumption is invalid. Even minor changes in user inputs or prompts produce distinct embeddings, making exact-match caching ineffective~\cite{Covington:2016:DNN:YouTube, gao2024retrievalaugmentedgenerationlargelanguage}. Consequently, ANN search systems effectively forfeit one of the most powerful tools in the systems toolbox.

% solutions so far
Existing ANN systems rely on index-internal heuristics—such as caching centroids~\cite{diskann} or upper graph layers~\cite{tiered-cache-hnsw}—to reduce average I/O. These mechanisms reduce constant factors but cannot eliminate backend execution, cannot exploit workload locality at the query level, and cannot generalize across ANN backends. Consequently, operators are forced into a persistent tradeoff between memory cost and latency, with no mechanism to amortize query work across time.

% insight
The key insight of this work is that query repetition in vector search is \emph{semantic} rather than exact. Real workloads exhibit temporal locality in embedding space: while query vectors are rarely identical, they are often sufficiently close that their nearest-neighbor sets are interchangeable at target recall. Exploiting this locality requires treating caching as a similarity problem rather than an exact lookup problem~\cite{similarity-caching}. However, similarity caching in high-dimensional vector spaces introduces two unresolved systems challenges: (1) selecting similarity thresholds that preserve recall without collapsing hit rates, and (2) performing cache lookups efficiently without turning the cache itself into a nearest neighbor index with unbounded latency and memory growth. Naively addressing either challenge leads to unbounded false positives that lower recall, or to cache designs whose overhead rivals the backend ANN system.

Prior work on similarity caching provides theoretical foundations but does not address the constraints of ANN serving systems~\cite{similarity-caching,similarity-caching-theory-algorithms}. There are existing practical systems for LLM response caching relying on a fixed global threshold~\cite{gptcache}, which perform poorly under prompts drawn from different distributions and therefore require careful retuning. %None provide a general, query-level cache that is backend-agnostic, recall-aware, and resource-bounded.

% our solution
We introduce \textbf{QVCache}, a new caching system for ANN search: a similarity-aware, recall-preserving query cache with fixed resource budgets. 
%Unlike traditional caches that map discrete keys to results, QVCache maps regions of the embedding space to recall-valid nearest-neighbor sets.
QVCache assigns region-specific similarity thresholds in the embedding space and dynamically adapts them using an online learning algorithm driven by observed nearest-neighbor distance statistics. This design bounds lookup latency and memory usage independently of dataset size, while aggressively short-circuiting backend ANN queries without compromising recall.

% impact of the solution

QVCache is designed as a drop-in layer for existing ANN systems. It requires no changes to index construction and applies uniformly to any backend vector database, whether in-memory or disk-based. Leveraging semantic locality, QVCache converts costly backend queries into cache hits, reducing both backend load and query latency with minimal memory overhead and without sacrificing recall. Instead of requiring large in-memory ANN indexes that grow with the dataset size, QVCache maintains much smaller in-memory mini-indexes whose size scales with the working set, which is typically far smaller in practice~\cite{incrementalivfindexmaintenance}.

% contributions
The contributions of this paper are:
\begin{itemize}[topsep=10pt, itemsep=0pt, parsep=0pt]
    \item A formulation of query-level vector caching as a similarity caching problem.

    \item The design of QVCache, a backend-agnostic ANN cache with adaptive, region-specific similarity thresholds and bounded latency and memory, which we show can reduce p50 latency by up to \(40\text{–}1000\times\) when integrated with existing ANN systems, while maintaining a memory footprint typically below 1\% of fully in-memory indexes.

    % \item An online threshold-learning algorithm driven by observed nearest-neighbor distance statistics.

    \item A workload generation benchmarking framework for controlled temporal-semantic locality in vector search, enabling reproducible evaluation of vector caching systems.

    \item Comprehensive experiments showing the benefits of QVCache under diverse datasets and backend vector databases (in-memory, disk-based, on-premise, or cloud-based).
\end{itemize}

\section{Background} 
This section reviews the foundations of vector search and similarity caching, and summarizes the FreshVamana index~\cite{freshdiskann}, which underpins the design of QVCache.
%In this section, we provide the preliminaries on vector search and the similarity caching problems. Further, we briefly review the FreshVamana index~\cite{freshdiskann}, which serves as the foundation for QVCache.

\textbf{Nearest Neighbor Search.} The \textit{k-nearest neighbor (k-NN)} problem, also referred to as vector search, can be formally defined as follows: Given a set of vectors $P$ in a $d$-dimensional space, i.e., $\forall p \in P, \; p \in \mathbb{R}^d$, and a query vector $q \in \mathbb{R}^d$, the objective is to return a set $X \subseteq P$ of $k$ vectors closest to $q$ according to a distance function $d$. That is, $|X| = k  \land \max_{x \in X} d(q, x) \le \min_{p \in P \setminus X} d(q, p)$. 
The exact solution relies on exhaustive search, which computes distances between $q$ and all vectors in $P$, incurring a time complexity of $O(|P| \cdot d)$.
%The most straightforward solution is \textit{exhaustive search}, which compares the query vector $q$ against all vectors in $P$, resulting in a time complexity of $O(|P|
% \cdot d)$. 
This cost is prohibitive at scale, motivating extensive research on \emph{Approximate Nearest Neighbor (ANN)} search. ANN methods trade exactness for efficiency by reducing the number of distance evaluations~\cite{diskann,HNSWMalkovY16,elpis,nsg,spann} and/or lowering the cost of individual distance computations~\cite{product-quantization}. This trade-off typically manifests as reduced recall.
%To reduce this computational cost, vector search research has largely focused on \textit{Approximate Nearest Neighbor (ANN) search}, which aims to minimize both the number of comparisons~\cite{diskann,HNSWMalkovY16,elpis,nsg,product-quantization,faiss,spann} and the cost of each comparison~\cite{product-quantization} with a trade-off of recall drop.

\textbf{Metrics.} The accuracy of ANN algorithms is commonly evaluated using \textit{k-recall@k}, defined as 
%To evaluate the accuracy of an ANN search method with respect to the ground-truth obtained via exhaustive search, the metric \textit{k-recall@k} is commonly used. It is computed as 
$
\frac{|X' \cap X|}{k},
$ 
where $X'$ is the result set returned by the ANN algorithm, $X$ represents the exact $k$-NN result obtained via exhaustive search, also known as ground-truth. System performance is typically characterized using \emph{latency} and \emph{throughput}, measured in queries per second (QPS).
%neighbors obtained from exhaustive search, and $k$ is the number of neighbors of interest for the query. For performance evaluation, \textit{latency} and \textit{throughput} (queries per second, QPS) are the two standard measures.

\textbf{Similarity Caching.}  
Vector caching builds upon the formal concept of the \textit{similarity caching} problem, where a user request for an object $O$ (in the vector caching setting, a set of top-$k$ nearest neighbor IDs) that is not in the cache may instead be served by a similar object $O'$ from the cache, at the cost of some degradation in user experience~\cite{similarity-caching-theory-algorithms}. The goal of similarity caching is to maximize the cache hit ratio while minimizing this degradation. A query $q$ requesting object $O$ results in a cache hit if there exists a cached object $O'$ retrieved by another query $p$ such that $d(q, p) \le r$, where $d$ is the distance (similarity) metric and $r$ is a similarity threshold~\cite{similarity-caching}. Choosing an appropriate value for $r$ is challenging: if $r=0$, the problem reduces to exact caching with nearly zero hit rate, while a large $r$ leads to poor user experience due to dissimilar results. Chierichetti et al.~\cite{similarity-caching} further show that the problem becomes even harder when $r$ is query-dependent, for which no competitive algorithm exists. Vector caching is exactly such a setting, where the appropriate similarity threshold is inherently query-dependent due to non-uniform distribution of database vectors in vector space, as discussed in Section~\ref{sec:threshold-challenges}. 
QVCache addresses this challenge by learning different similarity thresholds for different regions in the vector space via an online threshold learning algorithm (Algorithm~\ref{alg:learn-threshold}), in a supervised manner using feedback signals from the backend vector database.

\textbf{FreshVamana.}  \label{sec:background-freshvamana}
QVCache organizes cached vectors as multiple in-memory graph indexes, which we call \emph{mini-indexes}, and incrementally populates them on cache misses by inserting the vectors retrieved from the backend search. This setting requires a fully dynamic index that supports online insertions and efficient eviction. QVCache adopts the \emph{FreshVamana} algorithm~\cite{freshdiskann}, which is specifically designed for dynamic graph-based ANN indexing.
%QVCache manages vectors stored in memory as a graph structure. It starts empty and incrementally populates its graph index by fetching vectors from the backend database upon cache misses. Consequently, the in-memory index must be fully dynamic, supporting both insertions and deletions to enable efficient eviction. To achieve this, QVCache employs the \textit{FreshVamana} algorithm~\cite{freshdiskann}, which is specifically designed for dynamic graph construction.

Unlike static graph indexes, FreshVamana supports concurrent search and insertion. Query processing follows a greedy graph traversal, similar to other proximity graph methods. Upon inserting a new vector, the algorithm searches the existing graph to identify candidate neighbors and establishes edges to existing vectors accordingly. %To control graph degree and maintain navigability, FreshVamana applies a pruning procedure, \emph{RobustPrune}~\cite{freshdiskann}, whenever a node exceeds a predefined out-degree bound, removing redundant edges while preserving search quality.
%Unlike traditional static graph-based indexes, FreshVamana can build the index online. That is, it supports search operations concurrently with vector insertions, a property crucial for QVCache. Similar to other graph-based methods, it performs greedy search during query processing. When inserting a new vector, FreshVamana searches the existing graph to identify a set of candidate nodes, then connects the new vector to these candidates. To maintain graph sparsity, if any node’s out-degree exceeds a predefined threshold, a pruning procedure called \textit{RobustPrune}~\cite{freshdiskann} removes redundant edges while preserving navigability. FreshVamana also supports dynamic deletions; however, QVCache does not rely on this fine-grained deletion mechanism and instead adopts a coarser-grained eviction strategy through mini-indexes, tailored to its cache-oriented design.

Although FreshVamana supports fine-grained deletions, i.e. deletion of individual vectors, QVCache does not rely on this mechanism directly. Instead, it employs a coarser mini-index-level eviction policy, which reduces deletion overhead as discussed in Section~\ref{sec:eviction}.
\section{Solution Overview}
\begin{figure}[!htb]
  \centering

  % Top subfigure: cache hit, bottom part cropped away
  \begin{subfigure}[b]{0.75\linewidth}
    \centering
    \includegraphics[
      width=\linewidth,
      clip,
      trim=0 5cm 0 0 % left bottom right top (tune 3cm as needed)
    ]{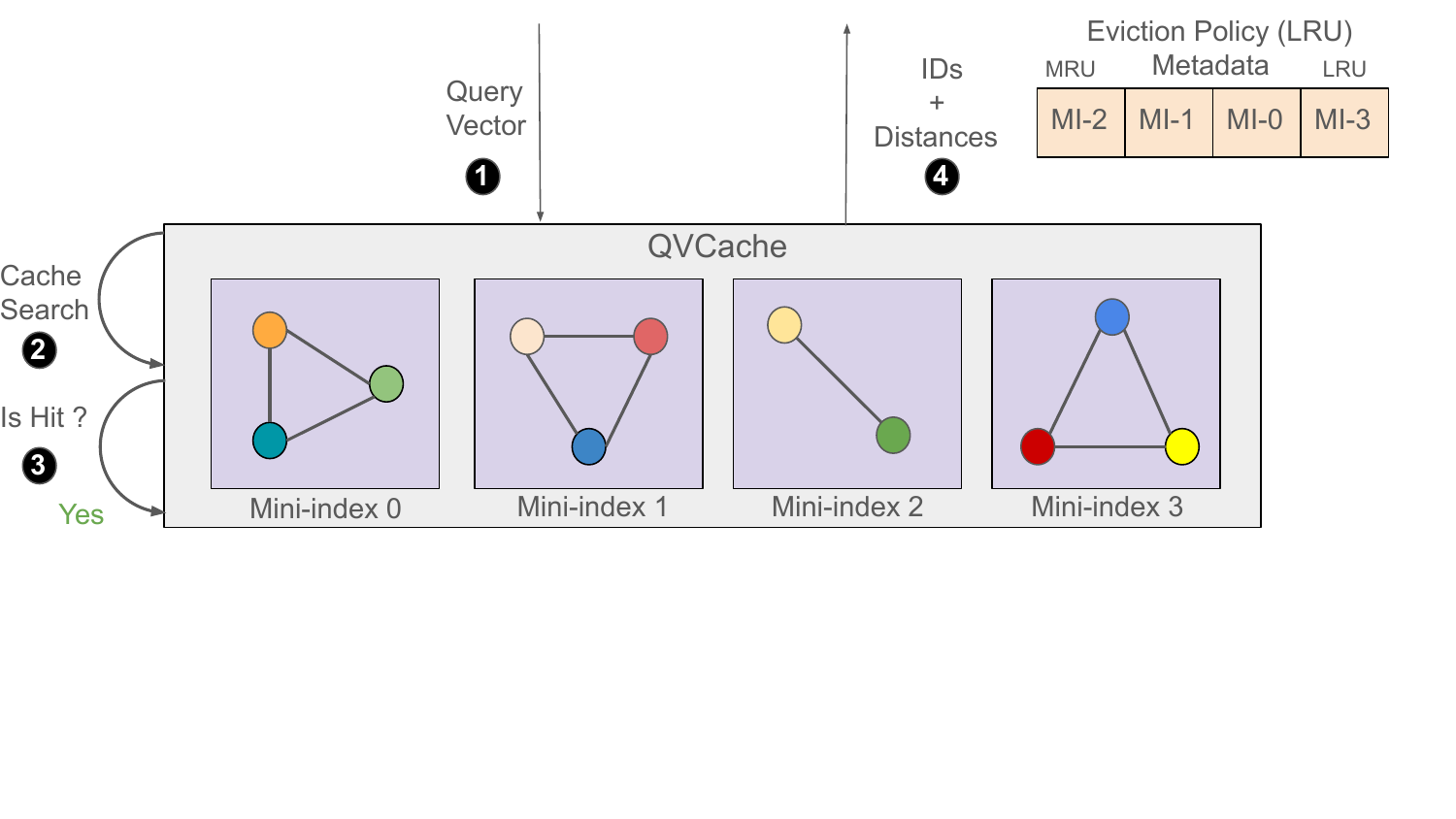}
    \caption{Cache hit}
    \label{fig:hit}
  \end{subfigure}
  
  \vspace{0.2cm} % <-- small vertical space between the two subfigures

  % Bottom subfigure: cache miss, unchanged
  \begin{subfigure}[b]{0.75\linewidth}
    \centering
    \includegraphics[width=\linewidth]{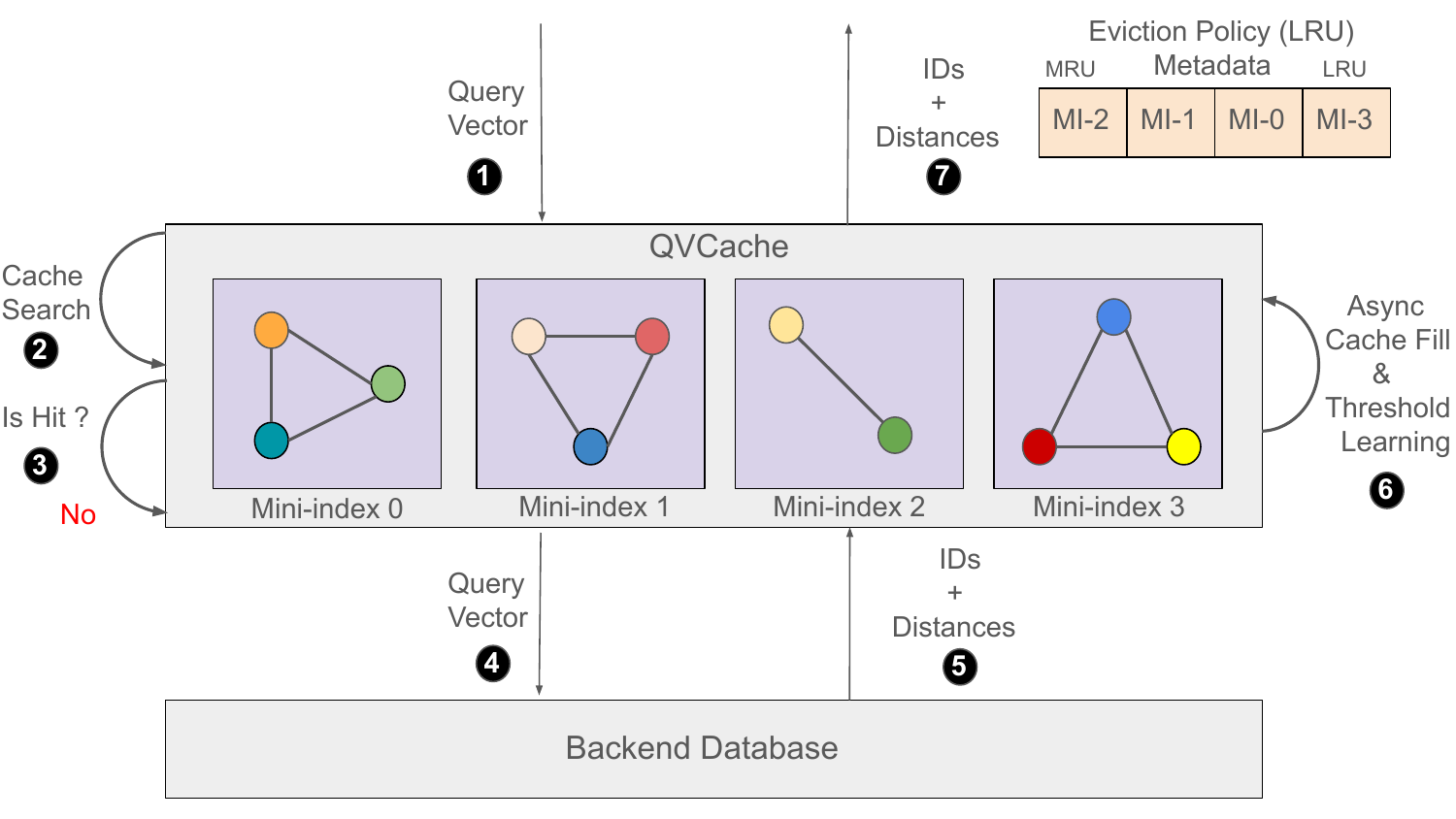}
    \caption{Cache miss}
    \label{fig:miss}
  \end{subfigure}

  \caption{Handling cache hits and misses in QVCache with four mini-indexes, each with a capacity of three vectors. Mini-indexes are ordered by the eviction policy metadata: the leftmost, MI-2, is the hottest (MRU), and the rightmost, MI-3, is the coldest (LRU).}
  \label{fig:architecture}
  \vspace{-\baselineskip}
\end{figure}

We introduce \textbf{QVCache}, a query-level cache for vector search that opportunistically bypasses backend ANN execution (Figure \ref{fig:hit}) when it detects that the query can be answered with high confidence (Algorithm \ref{alg:is-hit}) using cached vectors. QVCache is designed for workloads exhibiting \emph{temporal-semantic locality}, where queries recur within short time windows as nearby points in the embedding space, even though exact query vectors rarely repeat.
% Building on these observations, we introduce QVCache, the first (to our knowledge) query-level vector cache that bypasses the backend database when it can confidently answer a query using vectors previously retrieved on cache misses. QVCache is designed for workloads exhibiting temporal-semantic locality, where queries recur within a short time window and in semantically similar (i.e., spatially closer) forms. 

QVCache operates as a transparent layer in front of an existing vector database. It neither modifies backend index structures nor depends on backend-specific heuristics. Instead, it exploits workload locality to amortize expensive ANN execution across semantically similar queries, while bounding cache memory usage and lookup latency independently of dataset size.

\subsection{Workload Characteristics}  
\label{sec:workload-characteristics}
Traditional caching relies on temporal locality alone, assuming that identical requests recur. Vector search workloads violate this assumption: even minor variations in phrasing or prompts yield different embeddings. In e-commerce search, multiple query formulations can reflect the same purchase intent, and in conversational/RAG applications, users iteratively refine prompts within the same context. This motivates a similarity-based caching mechanism for workloads exhibiting \textit{temporal-semantic locality}, where receiving a query implies that semantically similar queries are likely to arrive in the near future.

Empirical studies across multiple domains show that such semantic repetition is widespread. In web search, about 30–40\% of queries are semantically repetitive variants of previous queries~\cite{10.1145/1277741.1277770yahoo, gill2025advancingsemanticcachingllms}; in recommender systems, this ratio rises to 55–68\%~\cite{li2024recommenderpurposerepeatexploration}; and in LLM-based RAG and conversational workloads, it can reach $\sim$70\%~\cite{gptcache, 10.1145/3721146.3721941}, where \cite{gptcache} defines repetition as pairs with cosine similarity at least 0.7 . Because these workloads ultimately issue queries via vector search, a comparable degree of semantic repetition is expected in the corresponding vector search queries as well.

Prior work further shows that temporal access patterns correlate with semantic similarity in real search workloads~\cite{semantic-similarity-temporal-correlation}; for example, seasonal effects in e-commerce cause temporally clustered queries such as “air conditioners” and “cooling systems” in summer, while semantically distant intents like “heaters” dominate in winter. 

Because semantically similar vector search queries tend to retrieve highly overlapping nearest-neighbor sets, caching vectors retrieved by one query can accelerate subsequent, nearby (semantically similar) queries without degrading recall.

Complementary empirical evidence from industrial systems shows that this reuse is highly concentrated, so the set of vectors worth caching is small. In large-scale Inverted File (IVF) indexes, only about 15\% of IVF partitions are accessed over an entire day~\cite{incrementalivfindexmaintenance}, and even within those hot partitions, only a small fraction of vectors are actually queried. As a result, the truly hot set over cache-relevant timescales (seconds to minutes) is often well below 1\% of the dataset. This pronounced access skew makes vector caching practical and effective. By maintaining a compact in-memory hot set, QVCache can serve cache hits with orders-of-magnitude lower latency than solely using disk-based backends, and even faster than fully in-memory ANN systems, while keeping memory usage strictly bounded at a level negligible compared to those systems.

%Unlike traditional caching, which relies solely on temporal locality, we coin the term \textit{temporal-semantic locality} to describe the behavior where semantically similar (i.e. spatially closer) queries repeat over time. This pattern is especially prevalent in e-commerce search~\cite{semanticequivalenceecommercequeries} and chatbot systems~\cite{gptcache}, where different phrasings reflect the same intent and yield nearby embeddings. Prior work further indicates that temporal access patterns correlate with semantic similarity in search workloads~\cite{semantic-similarity-temporal-correlation}. Because such queries often retrieve overlapping nearest-neighbor sets, QVCache can maintain a compact hot set of vectors in memory and efficiently serve repeated, similar queries.

%In an industrial vector search workload, production evidence shows that only about 15\% of IVF partitions are accessed over an entire day~\cite{incrementalivfindexmaintenance}. At the short timescales relevant for caching, for example minutes or even seconds, the active portion of the dataset is likely far below 1\%. This access skew makes vector caching highly practical. Leveraging these properties, QVCache achieves orders of magnitude lower latency (up to 300×) and corresponding throughput gains on a cache hit, while maintaining a minimal and bounded memory footprint.

\subsection{Query Flow through QVCache}  
QVCache processes every incoming query before it reaches the backend database. It first performs a \emph{cache search} over a collection of in-memory mini-indexes, each implemented as a small dynamic ANN graph i.e. a FreshVamana index. The cache search produces a candidate set of $k$ nearest neighbors.
%QVCache sits in front of a vector database of the user's choice, referred to as the \textit{backend database}. Incoming queries are first processed by QVCache, which searches through its in-memory mini-indexes. During this initial \textit{cache search}, QVCache generates a candidate set of $k$ nearest neighbors as specified by the query. 

To determine whether this candidate set is sufficiently accurate, QVCache compares the distance of the $k$-th neighbor against a \emph{region-specific similarity threshold}. If the distance falls within the threshold, the query is classified as a cache hit and answered directly from the cache, bypassing backend execution. These thresholds are not fixed globally; they are learned online from observed backend responses and adapt to local distance distributions in the embedding space, allowing QVCache to preserve recall while avoiding overly conservative cache decisions.
%If it determines that this candidate set is sufficiently accurate with respect to the unknown ground truth, the query is marked as a cache hit and answered directly from the cache, without contacting the backend database. The cache-hit decision is based on comparing the distance of the $k$-th neighbor in the candidate set to a region-specific \textit{similarity (distance) threshold}, which is dynamically learned over time from cache misses.

On a cache miss, the query is forwarded to the backend ANN system. Once the backend returns the resulting neighbor IDs, QVCache fetches the corresponding vectors from the backend and inserts them into the cache. Insertions are performed into the currently hottest mini-index, selected according to the cache metadata (most recently used mini-index). After cache population completes, QVCache updates the similarity thresholds using an online learning procedure driven by the newly observed nearest-neighbor distances.
%When a query results in a cache miss, it is redirected to the backend database for resolution. Upon receiving the response, QVCache populates its mini-indexes by fetching the corresponding $k$ vectors associated with the returned neighbor IDs. Afterwards, it updates the spatial thresholds using an online learning procedure. The fetched vectors are inserted into the hottest mini-index according to the eviction policy metadata (e.g., the most recently used mini-index in an LRU policy). 

Both vector fetching and threshold updates are performed asynchronously to keep the cache-miss latency on the critical path similar to the backend latency. This is essential in disk-based or remote deployments, where fetching vectors can incur multiple random I/O operations or network transfers. %To avoid using stale statistics, thresholds are updated only after cache insertion completes. As a result, vectors retrieved on a cache miss become available to future queries with a short delay, but never compromise recall.

%Both cache population and threshold updates are performed asynchronously to avoid increasing latency along the query's critical path, as fetching vectors from the backend can incur multiple random disk accesses ~\cite{turbocharging-vector-databases-ssds} or network transfers if QVCache is deployed on a separate machine. Threshold updates occur after the cache-fill completes to prevent using stale data, which could otherwise degrade recall. Consequently, vectors from a cache miss are not immediately available in the cache but become accessible with some delay.

QVCache initializes empty and fills its mini-indexes upon cache misses. Memory usage is bounded by construction: each mini-index has a fixed capacity, and eviction occurs at the granularity of entire mini-indexes rather than individual vectors. This design avoids fine-grained deletions, simplifies concurrency control, and ensures predictable memory overhead. The mini-index to evict is selected according to the configured policy, such as the least recently used mini-index, Mini-index 2, illustrated in Figure~\ref{fig:architecture}.
%To bound memory usage, each mini-index has an upper limit on the number of vectors it can store, and eviction is performed at the mini-index level rather than for individual vectors. The appropriate mini-index is evicted according to the configured policy metadata, such as the least recently used mini-index illustrated in Figure~\ref{fig:architecture}.

By bypassing backend ANN execution on cache hits, QVCache substantially reduces query latency and backend load. In cloud-based deployments where vector search is billed per query to the backend~\cite{ZillizServerless2025}, this reduction directly translates into lower operational cost, particularly at large scale.
%Thanks to this architecture, QVCache achieves a significant reduction in query latency by serving queries directly from the cache on a hit. The high cache-hit rate also translates into lower operational costs for cloud-based systems, where billing is often charged per query~\cite{ZillizServerless2025}, which can become expensive for large datasets.

In case of insertions and deletions of raw vectors other than the search, QVCache does not handle them directly but route them to the backend database, if the backend supports these operations. The resulting updates eventually propagate into QVCache through subsequent cache fills and evictions. %, and the approximate nature of vector search makes this eventual consistency acceptable in practice. 
This simplifies the cache management, and we leave supporting direct vector insertions and deletions as a future work, e.g., buffering them in QVCache and asynchronously applying them to the backend.
%As future work, we plan to add native support for insertions and deletions by buffering them in QVCache and asynchronously applying them to the backend, which we expect to further improve insertion and deletion throughput.
\section{Algorithms and Operations}
This section presents the overall architecture of QVCache, the unique challenges it addresses, and the algorithms and operations underlying its core components.

\subsection{Tiered Search}

\begin{algorithm}[htbp]
\caption{\textsc{TieredSearch}}
\label{alg:tiered-search}
\begin{algorithmic}[1]
\State \textbf{Input:} Query vector $Q$, result size $k$
\State \textbf{Output:} IDs of the $k$ nearest neighbors, their distances to $Q$
\State $(\text{ID}_{\text{cache}}, \text{d}_{\text{cache}}, \text{isHit}) \gets \textsc{CacheSearch}(Q, k)$ \label{alg:line:cachesearch}
\If{$\text{isHit}$} \label{alg:line:ishit}
    \State \Return $(\text{ID}_{\text{cache}}, \text{d}_{\text{cache}})$ \label{alg:line:returnhit}
\Else
    \State $(\text{ID}_{\text{backend}}, \text{d}_{\text{backend}}) \gets \textsc{BackendSearch}(Q, k)$ \label{alg:line:backendsearch}
    \State \textsc{AsyncCacheFill}$(\text{ID}_{\text{backend}})$ \label{alg:line:async-cache-fill}
    \State \textsc{AsyncLearnThreshold}$(Q, \text{d}_{\text{backend}}, k)$ \label{alg:line:async-learn-threshold}
    \State \Return $(\text{ID}_{\text{backend}}, \text{d}_{\text{backend}})$ \label{alg:line:returnmiss}
\EndIf
\end{algorithmic}
\end{algorithm}

\begin{algorithm}[htbp]
    \caption{\textsc{CacheSearch}}
    \label{alg:cache-search}
    \begin{algorithmic}[1]
    \State \textbf{Input:} Query vector $Q$, result size $k$, search strategy $\sigma \in \{\text{EAGER}, \text{EXHAUSTIVE}, \text{ADAPTIVE}\}$
    \State \textbf{Output:} IDs of candidate neighbors $\text{ID}_{\text{cache}}$, their distances $\text{d}_{\text{cache}}$ to $Q$, hit flag $\text{isHit}$
    \If{$\sigma = \text{ADAPTIVE}$} \label{alg:line:adaptive-check}
        \State $\text{hitRatio} \gets \textsc{GetHitRatioTrend}()$
        \State $\sigma \gets \begin{cases} \text{EXHAUSTIVE} & \text{if } \text{hitRatio} < \text{thresholdHitRatio} \\ \text{EAGER} & \text{otherwise} \end{cases}$ \label{alg:line:adaptive-switch}
        \State \Return $\textsc{CacheSearch}(Q, k, \sigma)$ \label{alg:line:return-adaptive}
    \Else \label{alg:line:eager-sequential}
        \State $R \gets \textsc{Reverse}(\textsc{GetEvictionOrder}())$ \label{alg:line:get-eviction-order}
        \State $(\text{ID}_c, \text{d}_c, \text{isHit}) \gets ([], [], \text{False})$ \Comment{candidates} \label{alg:line:init-candidates}
        \For{each mini-index $i \in R$} \label{alg:line:for-mini-indices}
            \State $(\text{ID}_i, \text{d}_i) \gets \textsc{SearchMiniIndex}(i, Q, k)$ \label{alg:line:search-mini-index}
            \If{$\textsc{IsHit}(Q, k, \text{d}_i)$} \label{alg:line:check-hit}
                \State $\text{ID}_c \gets \text{ID}_c.\text{concat}(\text{ID}_i)$, $\text{d}_c \gets \text{d}_c.\text{concat}(\text{d}_i)$ \label{alg:line:add-candidates}
                \State $\text{isHit} \gets \text{True}$ \label{alg:line:set-hit}
                \State $\textsc{UpdateMiniIndexEvictionMetadata}(i)$ \label{alg:line:update-lru}
                \If{$\sigma = \text{EAGER}$} \label{alg:line:check-eager}
                    \State \textbf{break} \label{alg:line:break-eager}
                \EndIf
            \EndIf
        \EndFor
        \State $(\text{ID}_c, \text{d}_c) \gets \textsc{ReRank}(\text{ID}_c, \text{d}_c)$ \label{alg:line:rerank}
        \State \Return $(\text{ID}_c[:k], \text{d}_c[:k], \text{isHit})$ \label{alg:line:return-results}
    \EndIf
    \end{algorithmic}
\end{algorithm}

QVCache is a query-level vector cache, analogous to systems like Redis~\cite{redis2025}, that operates in front of the main vector database (referred to as the backend database). Upon receiving a query, QVCache first attempts to answer it directly from the cache (Line \ref{alg:line:cachesearch} in Algorithm \ref{alg:tiered-search}). If it determines that the cached result is sufficiently reliable (Line \ref{alg:line:ishit}), the response is served without accessing the backend (Line \ref{alg:line:returnhit}). Otherwise, the query is forwarded to the backend database for processing (Lines \ref{alg:line:backendsearch}, \ref{alg:line:returnmiss}). In the background, asynchronous tasks fetch vectors from the backend and insert them into QVCache’s in-memory index structures, i.e., mini-indexes (Line \ref{alg:line:async-cache-fill}) and update the distance thresholds (Line \ref{alg:line:async-learn-threshold}) used in Algorithm \ref{alg:is-hit}. QVCache is \textit{backend-agnostic}, working with any vector database that provides the following interfaces, which most systems support:
\begin{itemize}
    \item \texttt{search(Q, k) → (ID[], d[])}: retrieves the nearest top-$k$ neighbor IDs and their distances to a query vector $Q$.
    \item \texttt{fetch(ID[]) → Vector[]}: fetches the vectors corresponding to the given IDs.
\end{itemize}

\begin{algorithm}[htbp]
\caption{\textsc{IsHit}}
\label{alg:is-hit}
\begin{algorithmic}[1]
\State \textbf{Input:} Query vector $Q$, result size $k$, distances of candidate neighbors to query $\text{d}_{\text{cache}}$ (sorted in ascending order)
\State \textbf{Output:} True for cache hit, False otherwise
\State $R \gets \textsc{ComputeRegionKey}(Q)$ \label{alg:line:compute-region-key} \
\If{$\text{d}_{\text{cache}}[k] \le (1 + D) \cdot \theta[k][R]$} \label{alg:line:cachecheck}
    \State \Return True
\EndIf
\State \Return False
\end{algorithmic}
\end{algorithm}

\subsection{Mini-indexes}
\label{sec:mini-indexes} QVCache maintains its in-memory vectors in multiple mini-indexes, each an instance of a FreshVamana graph~\cite{freshdiskann}, and manages them concurrently. This design narrows the search space during lookups (\text{EAGER} strategy in Algorithm~\ref{alg:cache-search}) and allows each mini-index to be treated as an independent unit for cache eviction.

\textbf{Cache Fill.}
QVCache maintains eviction metadata to track access patterns (i.e. cache hits, evictions, and fills) across its mini-indexes, ranking them from hottest to coldest based on the chosen eviction policy. Under the default LRU policy, the hottest mini-index corresponds to the most recently used one. Upon a cache miss, the system attempts to insert the new vector(s) into a mini-index (Line \ref{alg:line:async-cache-fill} in Algorithm \ref{alg:tiered-search}) in order of decreasing temperature, starting from the hottest and proceeding to progressively colder ones until it finds a mini-index with sufficient free capacity to accommodate all $k$ vectors, ensuring that all vectors from a single cache miss are colocated within the same mini-index rather than fragmented across multiple ones. If no mini-index has free capacity, QVCache evicts the coldest one according to the eviction policy, inserts the vector into that mini-index, and promotes it to be the hottest (similar to line \ref{alg:line:update-lru} in Algorithm \ref{alg:cache-search}). For example, in Figure~\ref{fig:architecture}, the system inserts into Mini-index~2, identified as the hottest according to the eviction policy.

\textbf{Cache Search.} In contrast to traditional caches with constant-time lookup, scanning a mini-index in QVCache (\textsc{SearchMiniIndex} in Algorithm~\ref{alg:cache-search}) incurs a time complexity of $O(\log c_{\text{mini-index}})$, where $c_{\text{mini-index}}$ denotes the capacity of each mini-index. Because QVCache maintains multiple mini-indexes, the worst-case lookup cost grows with both their number and size. Formally, let $C(N)$ denote the cost of searching a cache with capacity $N$ vectors,
partitioned into $n_{\text{mini-index}}$ mini-indexes, each storing $c_{\text{mini-index}}$ vectors.
The worst-case lookup cost then grows as

\begin{equation}
C(N) \propto n_{\text{mini-index}} \cdot \log\left(c_{\text{mini-index}}\right).
\label{exp:cache-search-cost}
\end{equation}

%Because nearest neighbors are colocated within a single mini-index during Cache Fill, searching only a small subset is sufficient.
To reduce this cost, QVCache tries to minimize the number of mini-indexes it scans. 
As in Cache Fill, it scans mini-indexes in descending heat order (Line~\ref{alg:line:get-eviction-order} in Algorithm~\ref{alg:cache-search}). Upon detecting a hit, QVCache inserts the corresponding neighbors into the candidate set (Lines~\ref{alg:line:check-hit}, \ref{alg:line:add-candidates}). Under the \texttt{EAGER} strategy, the search terminates after the first hit, whereas \texttt{EXHAUSTIVE} scans all mini-indexes and re-ranks the union of retrieved candidates (Line~\ref{alg:line:rerank}). Mini-indexes that contribute candidates to a cache hit are promoted to the hottest set (Line~\ref{alg:line:update-lru}). Because \texttt{EAGER} scans mini-indexes in the same heat order used during Cache Fill, and neighbors from the same cache miss are colocated in these hottest mini-indexes, it typically needs to scan only the first few mini-indexes to detect a cache hit (when one exists), effectively eliminating the linear factor in the cost expression in~\ref{exp:cache-search-cost}.

While \texttt{EXHAUSTIVE} strategy is more effective at recovering high-quality candidate sets, its cost grows linearly with $n{_\text{mini-index}}$. In contrast, \texttt{EAGER} exhibits near constant-time behavior with respect to $n{_\text{mini-index}}$ in typical settings; under high hit rates, the first mini-index alone often suffices. To balance these trade-offs, QVCache’s \texttt{ADAPTIVE} policy monitors recent hit ratios and switches between \texttt{EAGER} and \texttt{EXHAUSTIVE}: when the hit rate exceeds a threshold (e.g.,~0.9), it uses \texttt{EAGER}; otherwise, it falls back to \texttt{EXHAUSTIVE}. Empirically, \texttt{ADAPTIVE} performs well, though users may choose any strategy depending on their recall–latency requirements.

\vspace{1ex}

\textbf{Cache Eviction.}
\label{sec:eviction}
FreshVamana handles deletions in two stages. When a node is marked for deletion, it is first flagged as inactive and excluded from subsequent searches, but its memory remains allocated. Periodically, a background \textit{consolidation} process reclaims memory by removing all flagged nodes and reconnecting the surrounding graph. We observed that this lazy deletion strategy introduces two major drawbacks: (1) the consolidation step requires multi-threaded execution, adding significant overhead ~\cite{turbocharging-vector-databases-ssds}  and reducing throughput for workloads with frequently changing working sets ~\cite{working-set}, and (2) infrequent consolidation can cause memory bloat due to accumulated deleted nodes, which is problematic in memory-constrained environments where QVCache operates.

Because of these limitations, QVCache adopts mini-indexes as the unit of eviction. This enables to avoid costly consolidation operation in FreshVamana \cite{freshdiskann}. When all the mini-indexes get full, it evicts one of them according to eviction policy, e.g. Mini-index 3 in Figure~\ref{fig:architecture}, and promoted to the hottest index for the future insertions to be used. 

The larger the mini-indexes, the greater the information loss upon eviction: evicting a single mini-index removes a larger set of cached vectors at once. In the extreme case where $n_{\text{mini-index}} = 1$, a single eviction discards the entire cache. This motivates partitioning the cache into finer-grained units via multiple mini-indexes, thereby reducing eviction-induced information loss. However, increasing $n_{\text{mini-index}}$ also increases lookup latency through the linear factor in $n_{\text{mini-index}}$ in Equation~\ref{exp:cache-search-cost}. We study this capacity–partitioning trade-off empirically in Section~\ref{sec:granularity-matters}.

\vspace*{-\baselineskip} 

\subsection{Cache Hit and Miss Decisions}
\label{sec:threshold-challenges}
Deciding if a vector search query can be answered from cache (Algorithm~\ref{alg:is-hit}) is a similarity caching problem~\cite{similarity-caching, similarity-caching-theory-algorithms}. A query is considered a cache hit if the distance of the query vector to the furthest vector, $d_{\text{cache}}[k]$, in the candidate neighbor set is less than or equal to a \textit{similarity threshold}, $\theta[k][R]$, where $R$ is the identifier of the region the query falls into.. Determining the optimal threshold to maximize hit ratios while maintaining the recall of the backend database is challenging. If the threshold is too small, cache misses occur too frequently. If it is too large, cache hits return very different results from the backend, resulting in low recalls. We list four key challenges around the cache hit and miss decisions and respective solutions we propose.

\vspace{1ex}
 \textbf{Challenge 1: No universal threshold across datasets.} Different datasets may require substantially different distance thresholds, making manual tuning impractical.  

 \textit{Solution: Learned thresholds.} QVCache infers distance thresholds based on cache misses (Algorithm~\ref{alg:learn-threshold}), thereby automatically adapting these thresholds across different datasets.

\vspace{1ex}
 \textbf{Challenge 2: No universal threshold across data regions.} Vectors in a database are distributed non-uniformly across the high-dimensional space. Some regions are dense and highly clustered, while others are sparse. %, and certain regions contain more vectors than others.
 Consequently, a single global similarity threshold may perform well for some queries but poorly for others, even though the dataset is the same.

 \textit{Solution: Spatial thresholds.} It partitions the high-dimensional space into sub-regions (sub-spaces) and assigns a separate similarity threshold to each. These thresholds are learned independently, and for a given query, QVCache uses the threshold corresponding to the sub-region (sub-space) onto which the query vector falls to make cache hit decisions.

 %\begin{figure}[h]
 % \centering
 % \begin{minipage}[t]{0.48\linewidth}
 %   \centering
 %   \includegraphics[width=\linewidth]{figures/first.pdf}% first image
 % \end{minipage}
 % \hfill
 % \begin{minipage}[t]{0.48\linewidth}
 %   \centering
 %   \includegraphics[width=\linewidth]{figures/second.pdf}% second image
 % \end{minipage}
 % \caption{Initial queries help QVCache learn spatial similarity thresholds (left), and subsequent queries falling into these regions are classified as cache hits or misses (right). Blue dots represent data vectors, green dots queries resulting in cache hits, and red dots queries resulting in cache misses.}
 % \label{fig:spatial-thresholds}
%\end{figure}

\vspace{1ex}
    \textbf{Challenge 3: No universal thresholds over time.} Even within the same sub-region, query distribution may change, so the threshold that yields the best cache hit/miss decisions may change over time.
    
    \textit{Solution: Continuous learning.} QVCache runs Algorithm~\ref{alg:learn-threshold} continuously to adapt its thresholds, allowing it to track changes in query patterns.

\vspace{1ex}
 \textbf{Challenge 4: No universal threshold across different $k$ values.} As the parameter $k$ in a vector search query increases, the system generally requires a higher threshold. However, the threshold for a larger $k$ (e.g., $k=10$) cannot be inferred from that of a smaller $k$ (e.g., $k=1$), as the relationship is highly nonlinear and often unpredictable across datasets \cite{10.14778/3712221.3712224}.

 \textit{Solution: $k$-dependent thresholds.} QVCache maintains an independent threshold for each observed $k$ and learns them separately.

\vspace{1ex}
To implement these solutions, QVCache maintains a 2D array of thresholds, $\theta$. As shown in Algorithms ~\ref{alg:is-hit} and ~\ref{alg:learn-threshold} , the first dimension is resolved by the parameter $k$, while the second corresponds to the region $R$ into which the query falls, which is computed by \textsc{ComputeRegionKey} and described in detail in Section~\ref{sec:method:spatial-thresholding}. When evaluating whether a query is a cache hit, the system does not directly compare the distance of the furthest vector in the candidate set, $\text{d}_{\text{cache}}[k]$, with the threshold. Instead, it applies a multiplicative adjustment using the \textit{deviation factor}, $D$ (Line \ref{alg:line:cachecheck}). This factor serves as a tunable knob that balances cache hit ratio and recall: increasing $D$ raises the hit ratio but may lead to reduced recall. Such a knob provides flexibility for users to adapt QVCache to systems with different accuracy and performance requirements beyond the learned thresholds. In our experiments, we found that setting $D$ in the range $[0, 0.5]$ provides a practical operating regime.

\begin{algorithm}[htbp]
\caption{\textsc{LearnThreshold}}
\label{alg:learn-threshold}
\begin{algorithmic}[1]
\State \textbf{Input:} Query vector $Q$, result size $k$, distances of neighbors returned by the backend $\text{d}_{\text{backend}}$ (sorted in ascending order)
\State $R \gets \textsc{ComputeRegionKey}(Q)$
\State $\theta[k][R] \gets (1 - \alpha) \cdot \theta[k][R] + \alpha \cdot \text{d}_{\text{backend}}[k]$ \label{alg:line:updatetheta}
\end{algorithmic}
\end{algorithm}

  \vspace*{-\baselineskip} 
\subsection{Learning Thresholds}
In  Algorithm~\ref{alg:is-hit}, QVCache uses the distance of the furthest vector in the candidate set generated by itself, $\text{d}_{\text{cache}}[k]$, to determine the cache hit/miss. On the other hand, Algorithm~\ref{alg:learn-threshold} (called from Line \ref{alg:line:async-learn-threshold} of Algorithm \ref{alg:tiered-search}) uses the distance of the furthest vector returned by the backend, $\text{d}_{\text{backend}}[k]$ to learn the spatial thresholds in case of cache misses,  $\theta[k][R]$ in  Algorithm~\ref{alg:is-hit}. Ideally, $\text{d}_{\text{cache}}[k]$ converges to $\text{d}_{\text{backend}}[k]$ for a cache hit. 

The threshold update mechanism in QVCache was inspired by the Adam optimizer~\cite{adamoptimizer}, which updates momentum using gradients during neural network training. Similarly, QVCache updates thresholds using feedback from backend query results. The update employs an \textit{adaptivity rate}, $\alpha$, which determines how quickly QVCache adapts to changes in the query distribution (Line \ref{alg:line:updatetheta} in Algorithm \ref{alg:learn-threshold}). 

The first term in the update equation preserves the momentum of past query behavior, while the second term enables adaptation to query distribution shifts. A higher adaptivity rate, $\alpha$, allows QVCache to adjust more quickly to such shifts but also causes it to forget past query patterns faster. This balance enables QVCache to remain both stable and responsive under dynamic workloads.
In our experiments, we found that setting $\alpha = 0.9$ provides a good trade-off between stability and responsiveness.

%For each query resulting in a cache miss, QVCache populates the cache by fetching the top-$k$ vectors from the backend by their IDs and inserting them into the appropriate mini-indexes (Section \ref{sec:mini-indexes}), while also updating the thresholds using Algorithm~\ref{alg:learn-threshold}. To avoid using stale index data that could degrade recall, threshold updates are performed only after corresponding top-$k$ vectors are inserted into mini-indexes. Fetching vectors from the backend can be costly depending on the deployment strategy (i.e. whether QVCache and the backend are co-located or the backend is remote) and the backend's access path (i.e. whether it reads vectors from disk or keeps them in memory) to vectors. QVCache avoids blocking the critical query path, minimizing latency by executing both the cache and threshold updates asynchronously through a pool of background threads, as described in Algorithm~\ref{alg:tiered-search}.

%Due to the asynchronous execution of these tasks, if QVCache receives a query before the completion of cache update triggered by previous similar queries, it may result in a cache miss. This occurs because the thresholds used during this period are still stale, even though the query would have been a cache hit after the update is applied.

\vspace*{-\baselineskip} 
\subsection{Scalable Spatial Thresholding in High Dimensions}\label{sec:method:spatial-thresholding}
QVCache partitions the high-dimensional vector space into sub-spaces by dividing each dimension of the vector space into ${\text{n}}_{\text{buckets}}$ buckets. Upon receiving a query, it identifies the bucket corresponding to each dimension (\textsc{ComputeRegionKey}) to determine the appropriate spatial threshold.

However, pre-allocating the buckets for all possible sub-spaces is infeasible, especially for high-dimensional vectors.
For example, SIFT dataset vectors \cite{sift} have 128 dimensions, where storing the thresholds for all possible sub-spaces requires $8^{128}$ thresholds for ${\text{n}}_{\text{buckets}}$=8. 

\textbf{Dimensionality reduction.} Queries that are close in the original high-dimensional space remain close after projection into a lower-dimensional space, so the proximity relationships relevant for cache-hit decisions are approximately preserved. Therefore, QVCache projects incoming query vectors onto a lower-dimensional space via Principal Component Analysis (PCA) \cite{principalcomponentanalysis} and uses this reduced space for both threshold assignment and learning. QVCache performs a lightweight offline training step to compute the PCA transformation matrix, and this process does not require access to the full dataset; in our experiments, random sampling of as little as 0.1\%–0.01\% of the dataset was sufficient. On the SIFT dataset~\cite{sift}, this PCA training took around 10 minutes with a 0.1\% sampling ratio on the machine described in Section~\ref{sec:experimental-setup}, whereas building the corresponding ANN index took roughly a week. Thus, the additional training cost for QVCache is negligible compared to index construction.

On top of the projection, QVCache applies a straightforward partitioning scheme, dividing each dimension of the reduced space into equal-sized buckets. More advanced partitioning strategies that account for dataset-specific characteristics could be explored as future work. The \textsc{ComputeRegionKey} function applies the learned PCA transformation to project each query into the reduced space and identifies its corresponding region by locating the buckets it falls into.

For instance, if ${\text{d}}_{\text{reduced}} = 16$, the memory requirement becomes $8^{16}$ floating points for ${\text{n}}_{\text{buckets}}$=8. Although this technique significantly reduces the memory footprint, it still exceeds the practical constraints under which QVCache is designed to operate.

\textbf{Lazy initialization.} Similar to how data vectors tend to cluster, queries also exhibit spatial locality, meaning that not every region of the vector space will receive queries. Consequently, QVCache does not allocate memory for thresholds in regions that have not yet been queried; a missing key implicitly represents a region which has not been queried yet. Memory is allocated only when the first query for a new region arrives and the threshold is initialized to $d_{backend}[k]$ (i.e. updating the threshold by setting $\alpha$ to 1 in Algorithm \ref{alg:learn-threshold}). Our experiments show that, at most 50,000 regions (using the SpaceV  dataset \cite{SPACEV1B_SPTAG} with ${\text{d}}_{\text{reduced}} = 16$ and ${\text{n}}_{\text{buckets}}$=8), out of $8^{16}$ possible regions, become active, consuming approximately 200KB of memory. If the number of active regions exceeds a user-defined limit, QVCache can evict thresholds using an eviction policy such as LRU, and re-learn them later as needed. This lazy initialization, combined with dimensionality reduction, allows QVCache to maintain an exceptionally low memory footprint.

\begin{figure}[b]
    \vspace{-\baselineskip}
    \centering
    \includegraphics[width=0.75\linewidth]{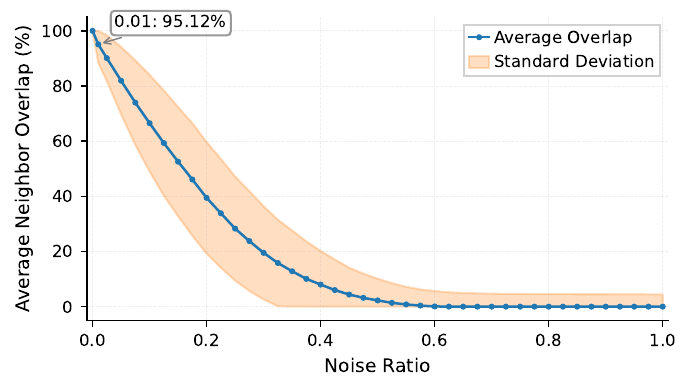}
    \caption{\textbf{Nearest Neighbor overlap under perturbation (DEEP \cite{deep} dataset, $k=10$).} The overlap between the neighbor sets of the base and perturbed queries decays sharply, approaching near-zero at a noise ratio of 0.5.}
    \label{fig:overlap-analysis}
\end{figure}

\begin{figure*}[t]
    \centering
    \begin{subfigure}[t]{0.48\textwidth}
        \centering
        \includegraphics[width=\linewidth,trim=0 120 0 0,clip]{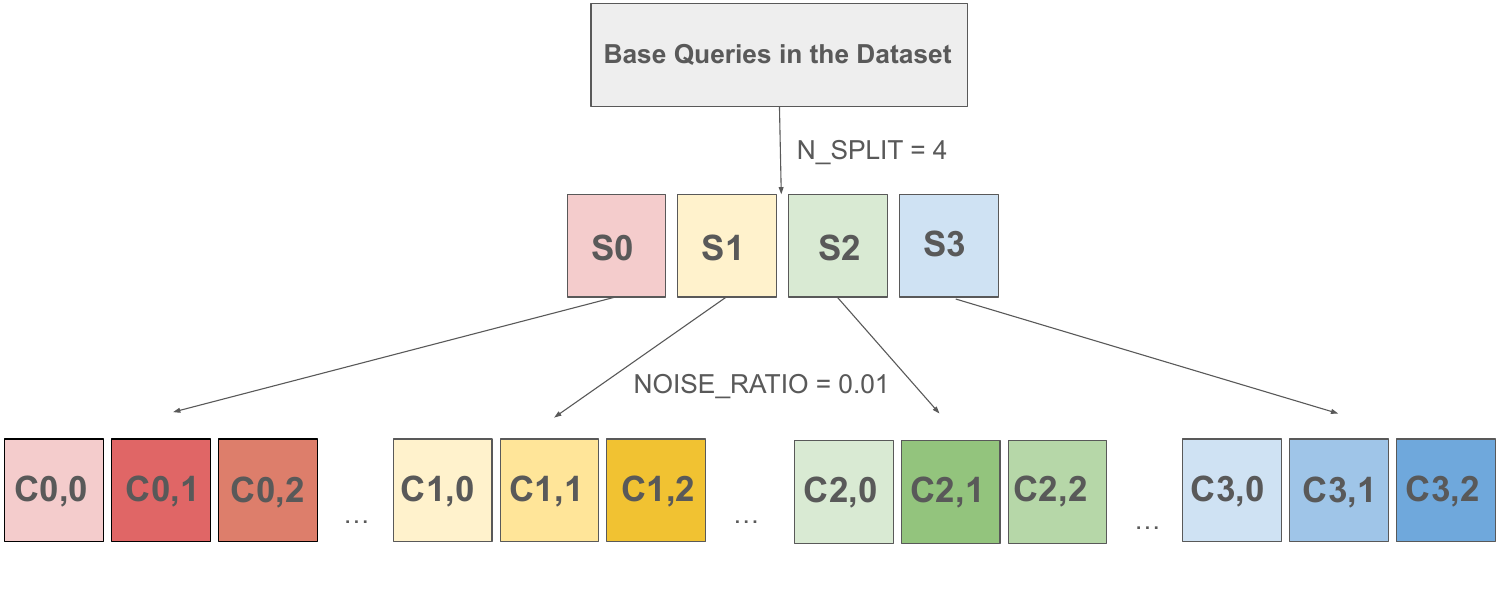}
        \caption{Synthesizing queries with semantic (spatial) locality.}
        \label{fig:query-perturbation}
    \end{subfigure}
    \hfill
    \begin{subfigure}[t]{0.48\textwidth}
        \centering
        \includegraphics[width=\linewidth,trim=0 120 0 0,clip]{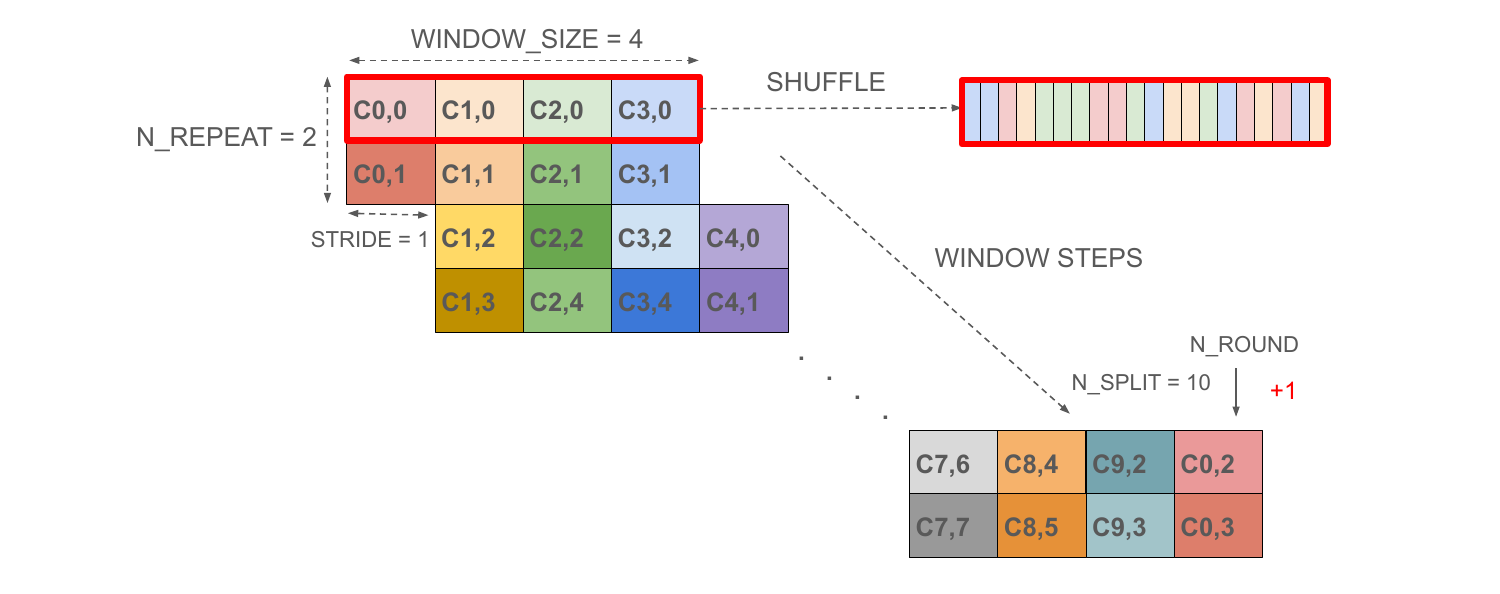}
        \caption{Workload generation with temporal–semantic locality.}
        \label{fig:sliding-window}
    \end{subfigure}
    \caption{Evaluation framework proposed in this paper for benchmarking vector caches, used to evaluate QVCache.}
    \label{fig:evaluation-framework}
    \vspace{-\baselineskip}
\end{figure*}

\section{Evaluation Framework}
%Real-world vector search workloads exhibit skewed access patterns \cite{298625SmartANN, incrementalivfindexmaintenance}, with spatially close queries recurring (i.e., {temporal-semantic locality}), but existing systems are typically evaluated without accounting for this behavior, which is inadequate for assessing vector caches. Standard benchmarks execute each query only once and report aggregate metrics such as average recall and latency, which is sufficient for evaluating standalone ANN systems but insufficient for understanding the cache behavior. 

Real-world vector search workloads exhibit skewed access patterns~\cite{298625SmartANN, incrementalivfindexmaintenance}, with spatially close queries recurring (i.e. \textit{temporal–semantic locality}). However, existing systems are typically evaluated without accounting for this behavior, which is inadequate for assessing vector caches. In standard benchmarks, queries provided in the datasets rarely have overlapping neighbors, and each query is executed only once; benchmarks then report aggregate metrics such as average recall and latency. While this is sufficient for evaluating standalone ANN systems, it is insufficient for understanding cache behavior due to the lack of temporal–semantic locality.

To address this, we propose a workload generation framework that produces query patterns exhibiting temporal-semantic locality at varying degrees. As illustrated in Figure \ref{fig:query-perturbation}, we first partition the queries from the dataset into $N_{\text{split}}$ disjoint subsets to model shifts in the working set of the workload. 
Within each subset (split), we generate perturbed variants for each query to model the temporal-semantic locality. For a query $q$, we sample a random vector $r$ from the data vectors in the dataset and produce  

\begin{equation}
q' = (1-\eta) \cdot q + \eta \cdot r
\label{exp:perturbation-expression}
\end{equation}

\noindent where $\eta$ controls the noise ratio. This interpolation yields queries that are semantically similar, i.e., spatially close, while remaining distinct. % enabling evaluation of vector caches. 
As shown in Figure \ref{fig:overlap-analysis}, the similarity between the base and perturbed queries' neighbor sets decreases sharply with increased noise. At $\eta = 0.01$, roughly 95\% of nearest neighbors overlap, simulating queries that differ in phrasing but share similar intent.

As visualized in Figure \ref{fig:sliding-window}, to generate recurrence of spatially close queries and drifts in the working set, we employ a windowed query pattern \cite{10.14778/2735461.2735465inmemoryperformanceforbigdata}. Each window consists of perturbed versions of $WINDOW\_SIZE$ many base splits. Queries within the window are randomly shuffled and dispatched to the system, repeating $N_{\text{repeat}}$ times. After each repetition, $stride$ out of $WINDOW\_SIZE$ perturbed splits are replaced with new ones. This process continues until the window reaches the last splits, and the cycle can optionally be repeated $N_{\text{round}}$ times with fresh perturbed copies.

The parameter $WINDOW\_SIZE$ controls the working set size (i.e., the number of vectors brought into the cache per window) of the workload. $N_{\text{repeat}}$ measures short-term memory, i.e., the ability to capture cache hits within a short time window, while $N_{\text{round}}$ measures long-term memory across multiple cycles. The ratio $stride/WINDOW\_SIZE$ adjusts how quickly the working set drifts \label{sec:drift-amount} \cite{10.14778/2735461.2735465inmemoryperformanceforbigdata}. Together, these parameters allow us to generate workloads with varying locality and temporal characteristics, enabling comprehensive evaluation of vector caches.

\section{Experiments}

In this section, we empirically investigate the following questions:

\begin{itemize}
    \item How well QVCache generalizes across datasets, query sizes ($k$), and workloads with varying degrees of temporal-semantic locality (Sections \ref{sec:dataset-experiments}, \ref{sec:temporal-semantic-locality-experiments}).
    
    \item What performance gains QVCache delivers when integrated with diverse backend systems (Sections \ref{sec:dataset-experiments}, \ref{sec:backend-agnostic-experiments}).

    \item How effective spatial thresholds are compared to a single global threshold (Section \ref{sec:spatial-thresholds}).
    
    \item How sensitive QVCache is to its hyperparameters 
    %(e.g., $D$, ${\text{d}}_{\text{reduced}}$, $n_{\text{mini-index}}$, $c_{\text{mini-index}}$)
    %and the impact on performance 
    (Sections \ref{sec:granularity-matters}, \ref{sec:cache-capacity-mini-index-partitioning},  \ref{sec:deviation-factor}, \ref{sec:space-partitioning}).
        
    \item What memory overhead QVCache incurs (Section \ref{sec:memory-overhead}).
\end{itemize}

\begin{table}[!htbp]
\centering
\small
\begin{tabular}{lrrrr}
\toprule
\textbf{Dataset} & \textbf{\#Vectors} & \textbf{Dim.} & \textbf{Distance} & \textbf{\#Queries} \\
\midrule
SIFT        & 1{,}000{,}000{,}000 & 128 & L2     & 10{,}000 \\
SpaceV\footnotemark[1] & 100{,}000{,}000     & 100 & L2     & 29{,}316 \\
DEEP\footnotemark[1]   & 10{,}000{,}000      & 96  & L2     & 10{,}000 \\
GIST        & 1{,}000{,}000       & 960 & L2     & 1{,}000 \\
GloVe       & 1{,}000{,}000       & 100 & Cosine & 10{,}000 \\
\bottomrule
\end{tabular}
\caption{Dataset statistics}
\label{tab:datasets}
\vspace{-\baselineskip}
\end{table}

\footnotetext[1]{For \textsc{SpaceV} and \textsc{DEEP}, we use the first 100M and 10M vectors of the 1B vectors in respective datasets.}

\subsection{Experimental Setup}
\label{sec:experimental-setup}
We implement QVCache in C++, together with Python bindings. All experiments are conducted on a Linux system in a containerized Docker environment, equipped with an Intel Xeon Gold 5118 processor, 2.30GHz, with 24 physical cores, 376GB of DDR4 RAM, and a Dell Express Flash PM1725a 1.6TB NVMe SSD.

\textbf{Datasets.} We evaluate how well QVCache generalizes across diverse data by benchmarking it on five datasets that differ in scale, domain, and dimensionality, as summarized in Table~\ref{tab:datasets} \cite{sift, deep, gist, pennington-etal-2014-glove, SPACEV1B_SPTAG}. 

\textbf{Backends.} We evaluate QVCache across a range of backend databases to understand its performance under diverse scenarios. We employ the DiskANN \cite{diskann} implementation by Yu et al. (2025) \cite{yu2025topologyawarelocalizedupdatestrategy}, a state-of-the-art disk-based vector search framework, for benchmarking QVCache across multiple datasets. Additionally, we test QVCache with FAISS \cite{faiss}, pgvector \cite{pgvector}, Qdrant \cite{qdrant2025}, Pinecone \cite{pinecone}, and SPANN \cite{spann} to assess its effectiveness with backends that differ in storage model (in-memory, disk-based, or hybrid), deployment model (on-premises vs. cloud-based), and index type (graph, tree, etc.).

\textbf{Metrics.} %We evaluate QVCache using six metrics, as illustrated in Figures \ref{fig:dataset-experiments} and \ref{fig:backend-experiments}. Metrics are collected at the granularity of window steps, with each dot representing a step. 
We evaluate QVCache across six metrics with values reported at window-step granularity, %where each point in the figures corresponds to a single step.
\emph{Cache hit ratio} measures the fraction of queries served by QVCache without forwarding requests to the backend database, while \emph{Hit latency} captures the latency of these queries. We use P50 latency and omit P99 latency because, unless QVCache achieves a hit ratio above 99\%, P99 is dominated by queries that miss the cache and are served by the backend. %yielding no meaningful distinction when using QVCache. %Instead, we use P50 latency to evaluate overall latency performance. %, independent of cache hits or misses.
Although QVCache is primarily designed for low-latency responses, it also improves throughput; we report the metric reflecting this benefit. To measure query accuracy, we use 10-recall@10. We also track the number of vectors retrieved from the backend into the cache over time to assess eviction behavior. 

\begin{figure}[h]
    \centering

    % Top row
    \begin{subfigure}[t]{0.48\linewidth}
        \centering
        \includegraphics[width=\linewidth]{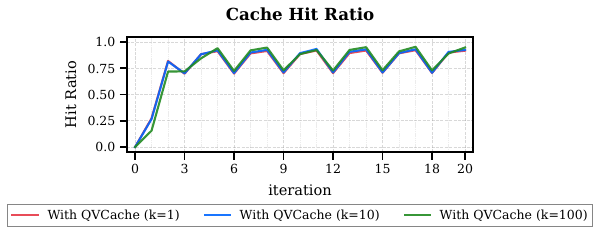}
    \end{subfigure}\hfill
    \begin{subfigure}[t]{0.48\linewidth}
        \centering
        \includegraphics[width=\linewidth]{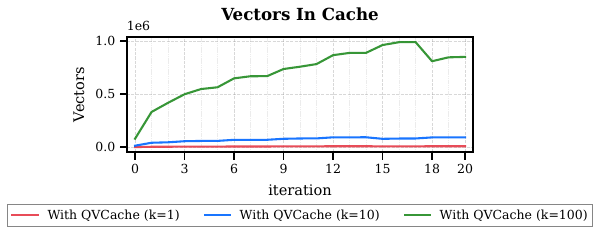}
    \end{subfigure}

    \vspace{0.75em}

    % Bottom row
    \begin{subfigure}[t]{0.48\linewidth}
        \centering
        \includegraphics[width=\linewidth]{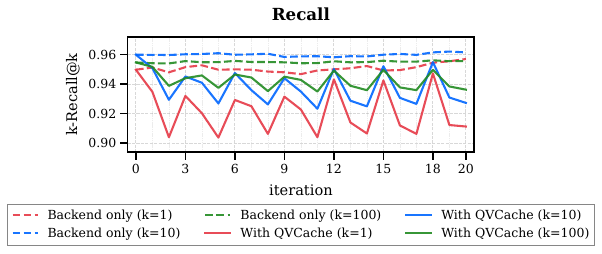}
    \end{subfigure}
    \begin{subfigure}[t]{0.48\linewidth}
        \centering
        \includegraphics[width=\linewidth]{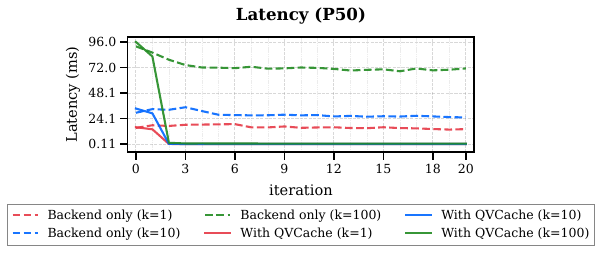}
    \end{subfigure}

    \caption{Effect of varying $k$ on the SIFT dataset. Cache capacity is 100{,}000 for $k = 10$ and is scaled linearly with $k$ (downscaled for $k = 1$ and upscaled for $k = 100$).}
    \label{fig:k-experiments}
    \vspace{-\baselineskip}
\end{figure}

\begin{figure*}[t]
\centering
% Define the column width for 5 columns (Using 0.195\textwidth for slight height increase)
\newlength{\colwidth}
\setlength{\colwidth}{0.195\textwidth}

% ===== Row 1: hit_ratio =====
\begin{subfigure}{\colwidth}
    \includegraphics[width=\linewidth]{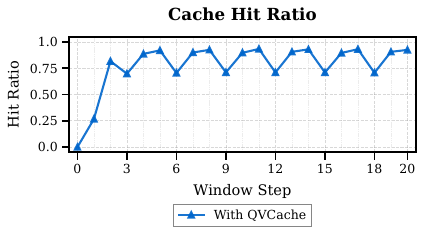}
\end{subfigure}
\begin{subfigure}{\colwidth}
    \includegraphics[width=\linewidth]{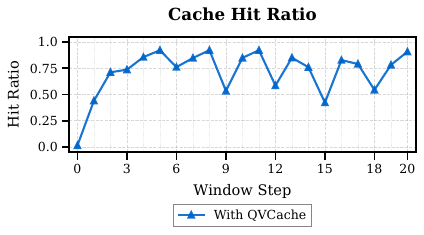}
\end{subfigure}
\begin{subfigure}{\colwidth}
    \includegraphics[width=\linewidth]{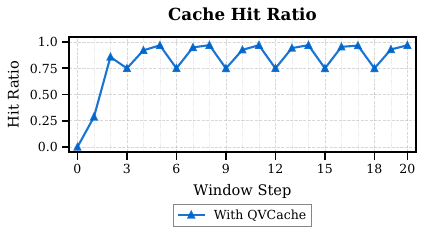}
\end{subfigure}
\begin{subfigure}{\colwidth}
    \includegraphics[width=\linewidth]{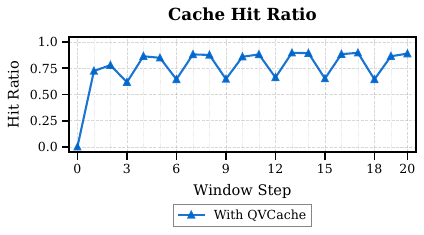}
\end{subfigure}
\begin{subfigure}{\colwidth}
    \includegraphics[width=\linewidth]{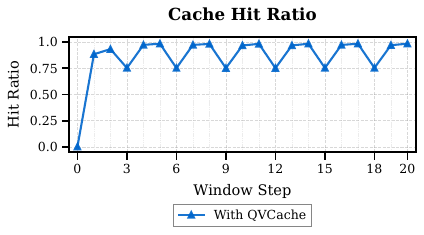}
\end{subfigure}
\\ % Add vertical space between rows

% ===== Row 2: avg_hit_latency (where the "shift" was observed) =====
\begin{subfigure}{\colwidth}
    \includegraphics[width=\linewidth]{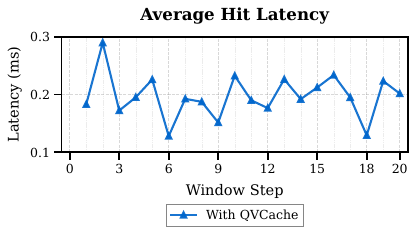}
\end{subfigure}
\begin{subfigure}{\colwidth}
    \includegraphics[width=\linewidth]{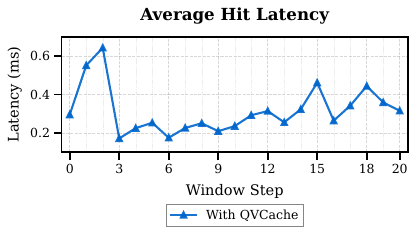}
\end{subfigure}
\begin{subfigure}{\colwidth}
    \includegraphics[width=\linewidth]{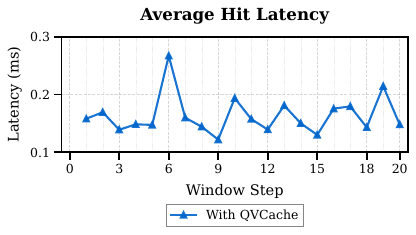}
\end{subfigure}
\begin{subfigure}{\colwidth}
    \includegraphics[width=\linewidth]{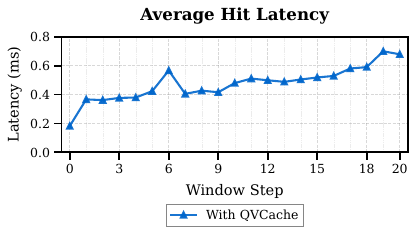}
\end{subfigure}
\begin{subfigure}{\colwidth}
    \includegraphics[width=\linewidth]{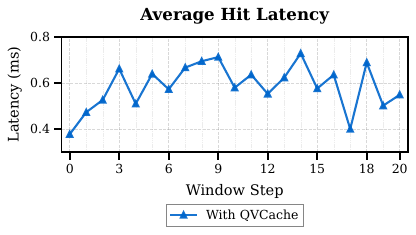}
\end{subfigure}
\\ % Add vertical space between rows

% ===== Row 3: p50_latency =====
\begin{subfigure}{\colwidth}
    \includegraphics[width=\linewidth]{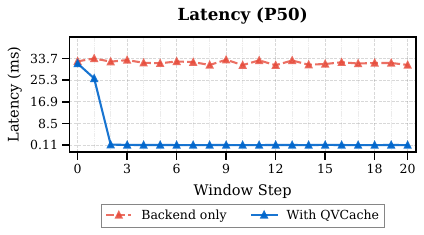}
\end{subfigure}
\begin{subfigure}{\colwidth}
    \includegraphics[width=\linewidth]{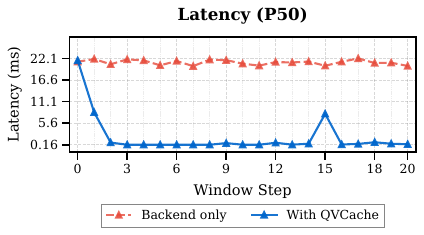}
\end{subfigure}
\begin{subfigure}{\colwidth}
    \includegraphics[width=\linewidth]{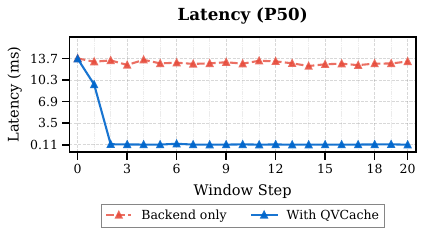}
\end{subfigure}
\begin{subfigure}{\colwidth}
    \includegraphics[width=\linewidth]{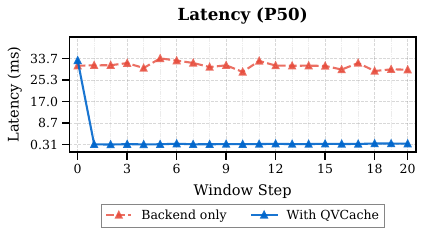}
\end{subfigure}
\begin{subfigure}{\colwidth}
    \includegraphics[width=\linewidth]{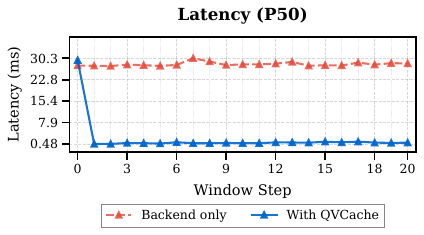}
\end{subfigure}
\\ % Add vertical space between rows

% ===== Row 4: qps =====
\begin{subfigure}{\colwidth}
    \includegraphics[width=\linewidth]{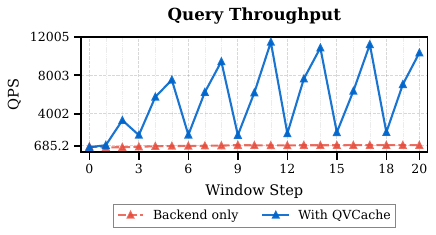}
\end{subfigure}
\begin{subfigure}{\colwidth}
    \includegraphics[width=\linewidth]{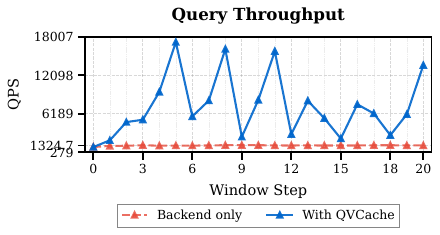}
\end{subfigure}
\begin{subfigure}{\colwidth}
    \includegraphics[width=\linewidth]{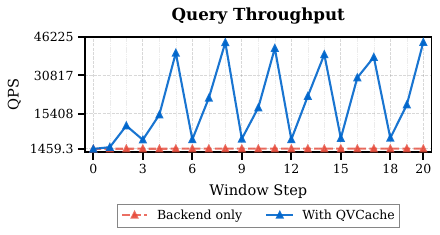}
\end{subfigure}
\begin{subfigure}{\colwidth}
    \includegraphics[width=\linewidth]{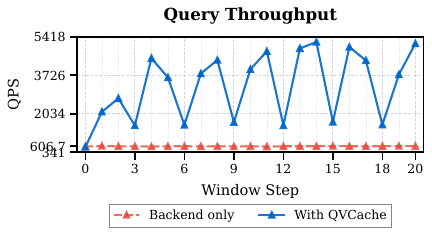}
\end{subfigure}
\begin{subfigure}{\colwidth}
    \includegraphics[width=\linewidth]{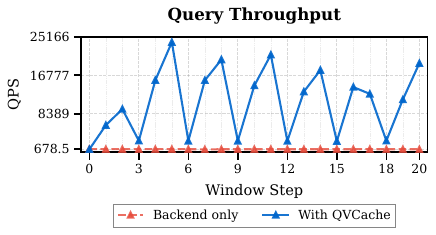}
\end{subfigure}
\\ % Add vertical space between rows

% ===== Row 5: recall =====
\begin{subfigure}{\colwidth}
    \includegraphics[width=\linewidth]{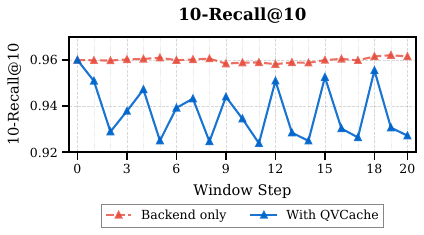}
\end{subfigure}
\begin{subfigure}{\colwidth}
    \includegraphics[width=\linewidth]{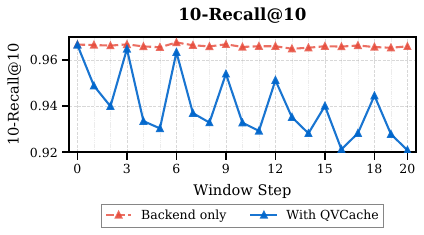}
\end{subfigure}
\begin{subfigure}{\colwidth}
    \includegraphics[width=\linewidth]{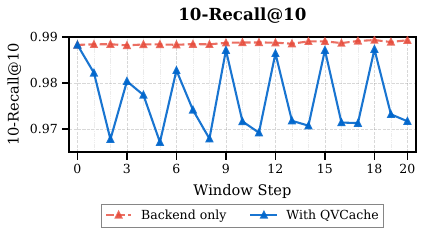}
\end{subfigure}
\begin{subfigure}{\colwidth}
    \includegraphics[width=\linewidth]{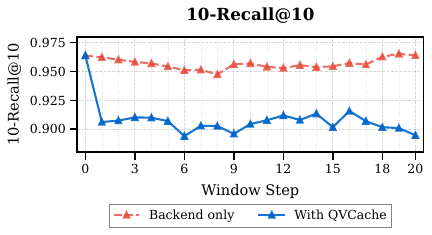}
\end{subfigure}
\begin{subfigure}{\colwidth}
    \includegraphics[width=\linewidth]{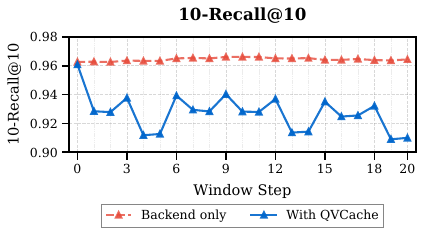}
\end{subfigure}

% ===== Row 6: Memory Active Vectors =====
\begin{subfigure}{\colwidth}
    \includegraphics[width=\linewidth]{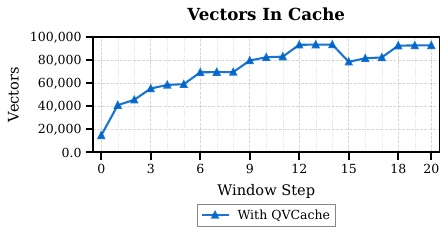}
\end{subfigure}
\begin{subfigure}{\colwidth}
    \includegraphics[width=\linewidth]{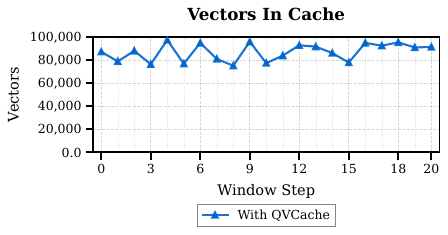}
\end{subfigure}
\begin{subfigure}{\colwidth}
    \includegraphics[width=\linewidth]{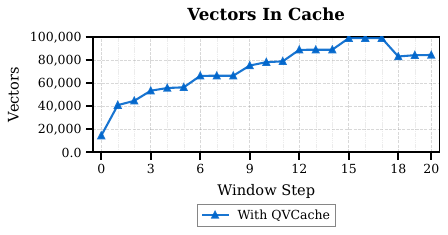}
\end{subfigure}
\begin{subfigure}{\colwidth}
    \includegraphics[width=\linewidth]{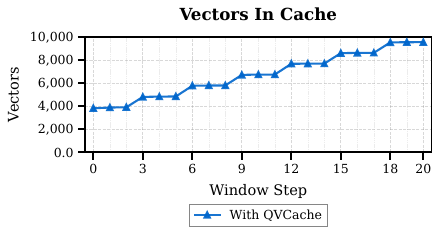}
\end{subfigure}
\begin{subfigure}{\colwidth}
    \includegraphics[width=\linewidth]{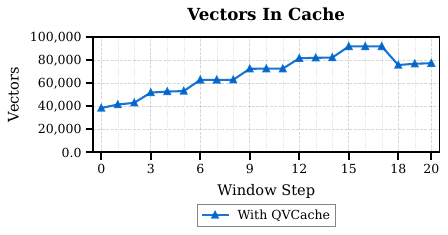}
\end{subfigure}

% ===== Row 7: Column captions (Datasets) =====
\begin{subfigure}{\colwidth}
    \centering
    \vspace{0.8em}
    (a) \textbf{SIFT}
\end{subfigure}
\begin{subfigure}{\colwidth}
    \centering
    \vspace{0.8em}
    (b) \textbf{SpaceV}
\end{subfigure}
\begin{subfigure}{\colwidth}
    \centering
    \vspace{0.8em}
    (c) \textbf{DEEP}
\end{subfigure}
\begin{subfigure}{\colwidth}
    \centering
    \vspace{0.8em}
    (d) \textbf{GIST}
\end{subfigure}
\begin{subfigure}{\colwidth}
    \centering
    \vspace{0.8em}
    (e) \textbf{GloVe}
\end{subfigure}

\caption{Vector search performance of backend vector database (DiskANN) alone vs. backend augmented with QVCache on the five datasets. $k$ is set to 10.}

\label{fig:dataset-experiments}
    \vspace{-\baselineskip}
\end{figure*}

\vspace*{-\baselineskip} 
\subsection{Adaptive Query-Aware Caching}
\label{sec:dataset-experiments}

A query-aware vector cache must adopt to non-stationary workloads, where the active working set drifts \cite{10.14778/2735461.2735465inmemoryperformanceforbigdata} over time. It should support varying dimensionalities, distance functions, and $k$ values while requiring minimal configuration (i.e., without manually tuning similarity thresholds), regardless of the underlying dataset.

The workloads in Figures \ref{fig:k-experiments} and \ref{fig:dataset-experiments} are generated with $N_{\text{split}} = 10$, $\eta = 0.01$, $N_{\text{repeat}} = 3$, $WINDOW\_SIZE = 4$, $stride = 1$, and $N_{\text{round}} = 1$ for the datasets in Table \ref{tab:datasets}, to empirically evaluate the query-awareness of QVCache.

QVCache is configured with $\alpha = 0.9$ (adaptivity rate), $n_{\text{buckets}} = 8$, and $d_{\text{reduced}} = 16$. The \texttt{ADAPTIVE} search strategy is applied. We set $D$ to 0.25 for SIFT and SpaceV, and to 0.075 for the remaining datasets. The cache capacity is fixed at 100{,}000 vectors and partitioned into $n_{\text{mini-index}} = 4$ mini-indexes, each with a capacity of 25{,}000, i.e. $c_{\text{mini-index}} = 25{,}000$ .

In Figure~\ref{fig:dataset-experiments}, when the working set stays stable for three consecutive window steps (i.e., the window shifts vertically in Figure~\ref{fig:sliding-window}), the hit ratio steadily increases as QVCache fills its mini-indexes with vectors from the active working set. Every third step, the working set changes by approximately 25\% (corresponding to a diagonal slide in Figure~\ref{fig:sliding-window}), which leads to a matching $\approx 25\%$ drop in hit ratio. %For example, with a stride of 2, this drop would be $\approx 50\%$. Throughput also exhibits a fluctuating pattern, driven by changes in the hit rate.

Queries that result in a cache hit have sub-millisecond latencies, typically between 0.1 and 1~ms. These high hit ratios, combined with sub-millisecond hit latencies, yield up to 60–300$\times$ lower p50 latency compared to using DiskANN only without QVCache.

Despite these significant performance gains, recall is only slightly impacted, dropping by 2–5\%. This impact can be further mitigated by tuning $D$, at the cost of some reduction in hit rate, as will be discussed in Section~\ref{sec:deviation-factor}.

\begin{figure*}[t]
\centering

% Adjusted column width for 5 columns (roughly 1/5 of text width)
\setlength{\colwidth}{0.18\textwidth}

% ===== Row 1: p50_latency =====
\begin{subfigure}{\colwidth}
    \includegraphics[width=\linewidth]{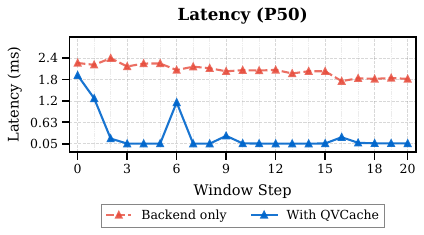}
\end{subfigure}
\begin{subfigure}{\colwidth}
    \includegraphics[width=\linewidth]{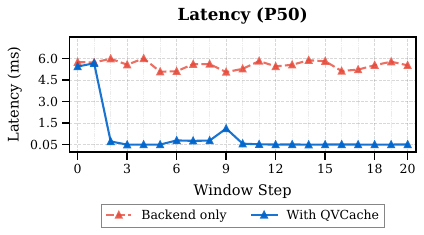}
\end{subfigure}
\begin{subfigure}{\colwidth}
    \includegraphics[width=\linewidth]{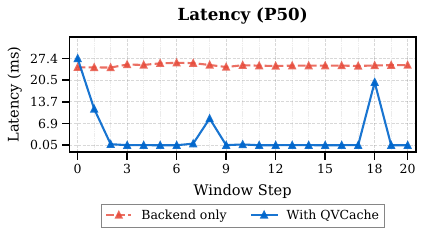}
\end{subfigure}
\begin{subfigure}{\colwidth}
    \includegraphics[width=\linewidth]{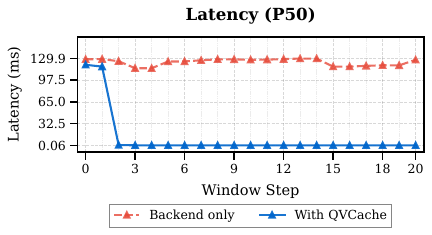}
\end{subfigure}
\begin{subfigure}{\colwidth}
    \includegraphics[width=\linewidth]{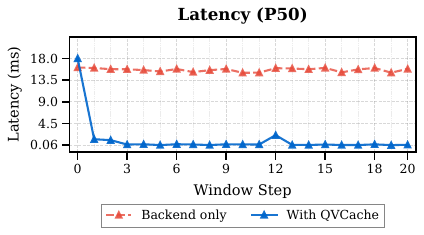}
\end{subfigure}

% ===== Row 2: recall =====
\begin{subfigure}{\colwidth}
    \includegraphics[width=\linewidth]{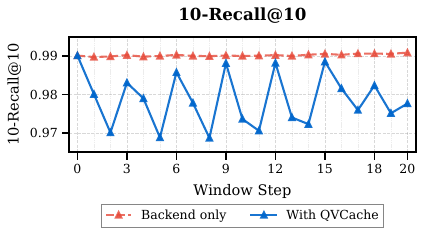}
\end{subfigure}
\begin{subfigure}{\colwidth}
    \includegraphics[width=\linewidth]{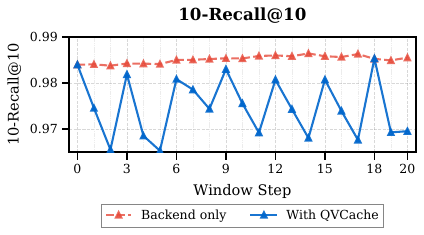}
\end{subfigure}
\begin{subfigure}{\colwidth}
    \includegraphics[width=\linewidth]{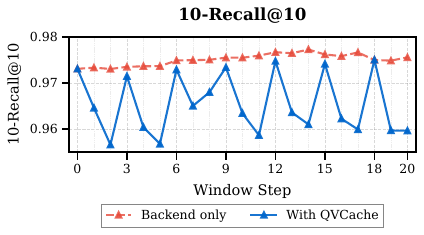}
\end{subfigure}
\begin{subfigure}{\colwidth}
    \includegraphics[width=\linewidth]{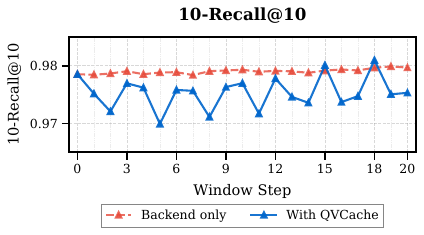}
\end{subfigure}
\begin{subfigure}{\colwidth}
    \includegraphics[width=\linewidth]{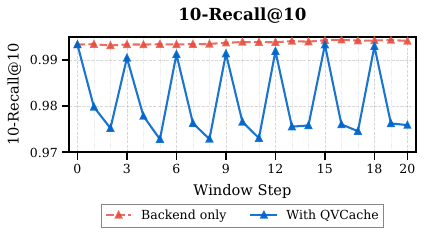}
\end{subfigure}

% ===== Row 3: Column captions =====
\begin{subfigure}{\colwidth}
    \centering
    \vspace{0.5em}
    \hspace{2em}(a) \textbf{FAISS}
\end{subfigure}
\begin{subfigure}{\colwidth}
    \centering
    \vspace{0.5em}
    \hspace{2em}(b) \textbf{Qdrant}
\end{subfigure}
\begin{subfigure}{\colwidth}
    \centering
    \vspace{0.5em}
    \hspace{2em}(c) \textbf{pgvector}
\end{subfigure}
\begin{subfigure}{\colwidth}
    \centering
    \vspace{0.5em}
    \hspace{2em}(d) \textbf{Pinecone}
\end{subfigure}
\begin{subfigure}{\colwidth}
    \centering
    \vspace{0.5em}
    \hspace{2em}(e) \textbf{SPANN}
\end{subfigure}

\caption{Performance of Various Backend Databases With and Without QVCache on DEEP Dataset}
\label{fig:backend-experiments}
    \vspace{-\baselineskip}
\end{figure*}

Moreover, QVCache adapts to varying $k$ values, as shown in Figure~\ref{fig:k-experiments}. Across all $k$ values, QVCache preserves high recall and hit ratio, while latency reductions become more pronounced with larger $k$, reaching up to 950$\times$ lower for $k = 100$.

\subsection{Evaluating QVCache with Different Backends}
\label{sec:backend-agnostic-experiments}
QVCache is compatible with any vector search system (i.e. backend-agnostic), independent of the underlying index type, system scale, or deployment environment, requiring only the implementation of standard search and fetch interfaces. To quantify this property, we repeat the experiment from Section~\ref{sec:dataset-experiments} on DEEP dataset with different backends, as shown in Figure~\ref{fig:backend-experiments}. All backends are evaluated both with and without QVCache. FAISS, Qdrant, SPANN and pgvector are deployed on the same Linux host, while Pinecone is evaluated using its managed cloud service.

Pinecone \cite{pinecone}, a cloud-managed vector search service, exhibits relatively high latencies ($\approx$ 100 ms) due to network round-trip overheads. As shown in Figure \ref{fig:backend-experiments}, integrating QVCache on the client side bypasses this network latency and reduces p50 latency by up to three orders of magnitude ($\approx$ 1000$\times$). Although cache-miss fetches may incur additional time, we observe no degradation in recall or hit-rate convergence. %Beyond latency reduction, QVCache also lowers serving costs for services billed on a per-query basis by converting a large fraction of requests into local, non-billable cache hits.

For hybrid memory–disk backends, such as Qdrant \cite{qdrant2025} and SPANN \cite{spann}, and disk-only backends like pgvector \cite{pgvector}, QVCache provides substantial latency improvements by fronting their client libraries. Specifically, we observe up to $\approx$100$\times$, $\approx$300$\times$, and $\approx$500$\times$ reductions in p50 latency for Qdrant, SPANN, and pgvector, respectively. While our experiments use client-side integration, embedding QVCache directly within these systems would enable cross-client caching, exploiting a global view of incoming queries and allowing multiple clients’ requests to be served more efficiently through better aggregation, which would be an interesting future work.

Even for fully in-memory backends such as FAISS, QVCache achieves up to 40$\times$ latency reduction. Here, both the backend and QVCache maintain indexes and vectors in memory, yet cache hits are faster because QVCache constrains the search space, allowing best-first search to converge in fewer steps. %Nonetheless, QVCache introduces an in-cache probe for every request; when backend latency is already low and cache hit rates are limited, this overhead can be non-negligible.

%In summary, augmenting a state-of-the-art in-memory backend such as FAISS with QVCache results in approximately $40$$\times$ lower p50 latency. For disk-based vector search systems, augmenting them with QVCache yields $\approx 100$–$ 500$$\times$ reductions in p50 latency, and for cloud-hosted systems such as Pinecone, the benefit is magnified to nearly $1000$$\times$ by eliminating network latency on cache hits.

\subsection{Spatial Thresholds vs. Global Threshold}
\label{sec:spatial-thresholds}
Vector distributions vary significantly across the vector space: some regions are densely clustered, while others are sparse. This heterogeneity makes it impractical to rely on a single global similarity threshold for all cache hit decisions.

\begin{figure}[t]
    \centering

    \begin{subfigure}{0.7\linewidth}
        \centering
        \includegraphics[width=\linewidth]{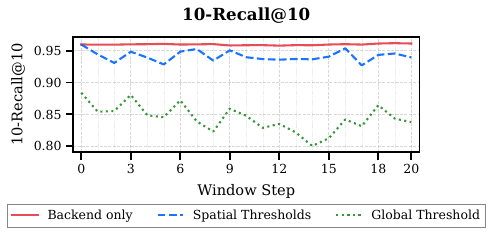}
        \label{fig:plot1}
    \end{subfigure}

    \caption{Impact of Global vs. Spatial Threshold(s) on Recall on SIFT}
    \label{fig:spatial-threshold-experiment}
    \vspace{-\baselineskip}
\end{figure}

To evaluate this effect, we repeated the experiment from Section \ref{sec:dataset-experiments} on the SIFT dataset under two configurations: one using a single global threshold and the other using spatial thresholds. As shown in Figure~\ref{fig:spatial-threshold-experiment}, using spatial thresholds preserves recall with at most a 2–3\% drop, whereas a single global threshold can incur losses of up to 16\%. The underlying reason is that spatial thresholds learn locally appropriate hit/miss sensitivities, while a single global threshold fails to capture local variations and thus degrades recall.

\subsection{Granularity Matters: Balancing Eviction Cost and Hit Latency}
\label{sec:granularity-matters}
As discussed in Section~\ref{sec:mini-indexes}, given a fixed cache capacity, we can reduce eviction-induced information loss by partitioning the cache across multiple mini-indexes. However, increasing the number of mini-indexes worsens cache lookup and consequently hit latencies, as indicated by the cost expression in Equation~\ref{exp:cache-search-cost}.

\begin{figure}[h]
    \centering
    \begin{subfigure}{0.48\linewidth}
        \centering
        \includegraphics[width=\linewidth]{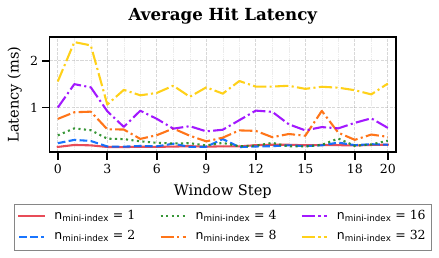}
        \label{fig:avg-hit-latency}
    \end{subfigure}
    \hfill
    \begin{subfigure}{0.48\linewidth}
        \centering
        \includegraphics[width=\linewidth]{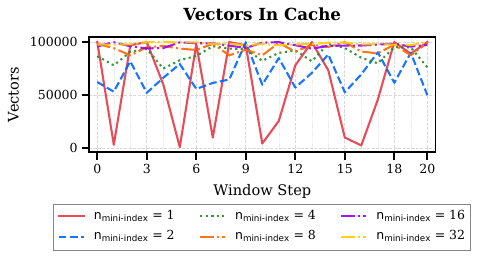}
        \label{fig:memory-active-vectors}
    \end{subfigure}

    \caption{Effect of mini-index granularity on QVCache performance on the SpaceV dataset. The total cache capacity is fixed at 100{,}000 vectors, while the number of mini-indexes is varied to control per–mini-index capacity.}
    \label{fig:granularity-effect}
\end{figure}

To further analyze this trade-off, we conducted the experiment shown in Figure~\ref{fig:granularity-effect} using the SpaceV dataset. We fixed the total cache capacity and varied the number of mini-indexes, $n_{\text{mini-index}}$, which implicitly determines the capacity of each mini-index, $c_{\text{mini-index}}$. As $n_{\text{mini-index}}$ increases, the average cache-hit latency rises, since the scanning strategy must probe a larger number of mini-indexes before identifying a confident candidate neighbor set (Equation~\ref{exp:cache-search-cost}), even under the \texttt{EAGER} strategy. Conversely, eviction cost increases as $n_{\text{mini-index}}$ decreases, as evidenced by the sharp drops in cache vectors for $n_{\text{mini-index}}=1$ and $n_{\text{mini-index}}=2$ in Figure~\ref{fig:granularity-effect}, indicating large, bursty eviction events and increased eviction-induced information loss. Accordingly, we choose $n_{\text{mini-index}} = 4$ as a balanced operating point between eviction-induced information loss and cache-hit latency.

%For scaling the cache capacity, we recommend first estimating the working set size, $w$, of the workload and setting the capacity of each mini-index to this value, that is, $c_{\text{mini-index}} = w$. This choice allows the \texttt{EAGER} strategy to terminate after scanning only the first one or two mini-indexes in most cases. Therefore, it avoids over-partitioning by eliminating the linear dependence on $n_{\text{mini-index}}$ in Equation~\ref{exp:cache-search-cost}, replacing it with an approximately constant multiplier $c \approx 1\text{--}2$. The total cache capacity can then be scaled to $N$ vectors by adding additional mini-indexes of size $w$. As a result, the cache lookup cost becomes nearly constant with respect to the cache size $N$ and is well-approximated by $c \log w$. Moreover, as the working set changes, vectors from previous working sets become cold and are naturally evicted at the granularity of mini-indexes, which further reduces eviction-induced information loss.

\begin{figure}[h]
\centering
\setlength{\colwidth}{0.48\columnwidth}

% ===== Column 1 =====
\begin{minipage}[t]{\colwidth}
    \centering
    \includegraphics[width=\linewidth]{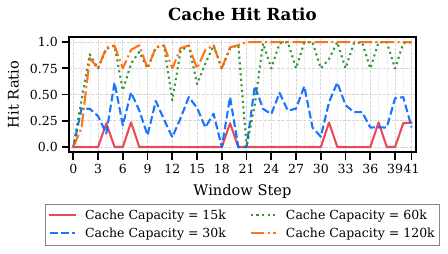}\\[0.5em]
    \includegraphics[width=\linewidth]{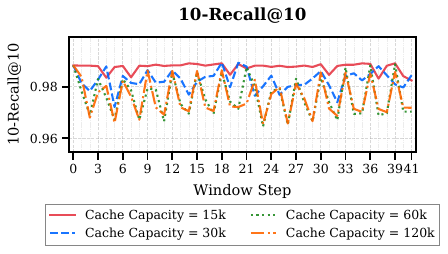}\\[0.5em]
    \includegraphics[width=\linewidth]{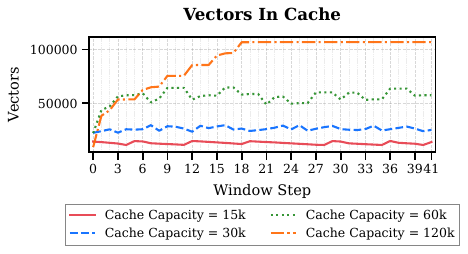}\\[0.5em]
    \vfill
    \textbf{(a) Varying Cache Capacity}
    \label{fig:cache_size_experiments:a}
\end{minipage}
\hfill
% ===== Column 2 =====
\begin{minipage}[t]{\colwidth}
    \centering
    \includegraphics[width=\linewidth]{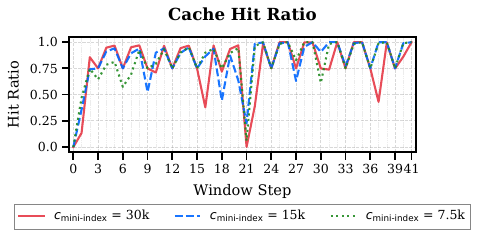}\\[0.5em]
    \includegraphics[width=\linewidth]{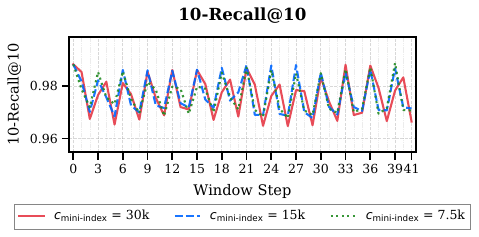}\\[0.5em]
    \includegraphics[width=\linewidth]{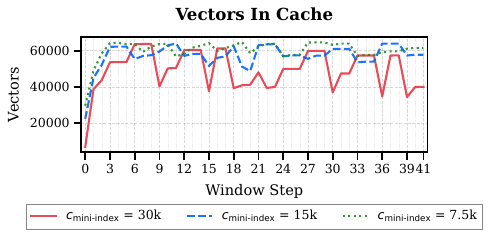}\\[0.5em]
    \vfill
    \textbf{(b) Varying Mini-index capacity}
    \label{fig:cache_size_experiments:b}
\end{minipage}

\caption{
Effect of cache capacity and size of mini-indexes in QVCache on the SIFT dataset.
Left: varying total cache capacity via $n_{\text{mini-index}}$ with fixed $c_{\text{mini-index}}$.
Right: varying $c_{\text{mini-index}}$ with fixed total capacity.
}
\label{fig:cache_size_experiments}
\end{figure}

\subsection{Sensitivity to Cache Capacity and Mini Index Partitioning}
\label{sec:cache-capacity-mini-index-partitioning}
%As first noted by Belady~\cite{belady1966study}, an ideal cache would retain exactly the items that will be accessed again and evict only those that will never be referenced in the future. Such optimal behavior is attainable only with an unbounded cache or with an oracle that perfectly predicts future accesses. Since neither assumption is practical, real systems are inherently constrained by finite capacity and imperfect eviction decisions. The same limitations apply to QVCache.

To see how well QVCache captures cache hits under varying cache capacities and varying mini-index sizes, we repeat the experiment from Figure~\ref{fig:dataset-experiments} on the SIFT dataset with $N_{\text{round}} = 2$ as shown in Figure~\ref{fig:cache_size_experiments}.

For the generated workload we have $4$ perturbed copies ($C_{i,j}$'s in Figure \ref{fig:evaluation-framework}) of 4 split ($S_i$'s in Figure \ref{fig:query-perturbation}) in each window since $WINDOW\_SIZE = 4$. Each perturbed copy of a split, brings approximately $15{,}000$ vectors into the cache. Therefore, the working set size of the workload, $w$, becomes $60{,}000$. Since $\textit{stride} = 1$, approximately one quarter of the working set changes at each window slide (every $N_{repeat} =3$ window steps).

Similar to any other cache, if its capacity, $N$, is not large enough to fit the working set, the hit ratio drops due to frequent evictions. We observe the same effect for QVCache in Figure ~\ref{fig:cache_size_experiments}a, where the cache capacity is varied while the mini-index capacity $c_{\text{mini-index}}$ is fixed at $15{,}000$. We see that when the cache capacity is smaller than the working set size, the hit ratios (red and blue lines) drop severely, as expected, whereas they remain high when the capacity is greater than or equal to the working set size (green and orange lines), also as expected. Moreover, while $N = 60{,}000$ shows a drop in hit ratio when the second round starts (at window step $21$), it stays constant at $1$ for $N = 120{,}000$, because the $N = 120{,}000$ setting is able to keep the vectors fetched from the first round in the cache, whereas the $N = 60{,}000$ setting has already evicted them and therefore has to bring them into memory again. Although we observe hit-ratio drops in the insufficient-capacity settings, recall remains unaffected and is in fact higher, since the majority of queries are then answered directly by the backend database.

We then fixed the total cache capacity at $N = 60{,}000$, which is just sufficient to hold the working set, and varied the mini-index capacity $c_{\text{mini-index}}$. In Figure~\ref{fig:cache_size_experiments}b, we see that QVCache is not sensitive to $c_{\text{mini-index}}$ in terms of hit ratio and recall, with the exception that eviction-induced information loss increases as $c_{\text{mini-index}}$ grows. %However, the number of vectors stored in the cache (red line) drops (eviction-induced information loss; see Section~\ref{sec:mini-indexes}) much more sharply as $c_{\text{mini-index}}$ increases.

Therefore, we recommend setting the total cache capacity $N$ (in vectors) large enough to accommodate the expected working set size, as is standard practice for any caching system, and generally larger. Due to the access skew described in Section~\ref{sec:workload-characteristics}, this is relatively inexpensive to provision. The working set size grows linearly with $k$ and with the number of diverse (semantically dissimilar) queries arriving within a reuse interval which is proportional to $WINDOW\_SIZE$ in our experiment.

For partitioning the cache capacity, we recommend first estimating the working set size, $w$, of the workload in a best-effort manner and setting the capacity of each mini-index to this value, that is, $c_{\text{mini-index}} = w$. This choice allows the \texttt{EAGER} strategy to terminate after scanning only the first one or two mini-indexes in most cases. Therefore, it avoids over-partitioning by eliminating the linear dependence on $n_{\text{mini-index}}$ in Equation~\ref{exp:cache-search-cost}, replacing it with an approximately constant multiplier $c \approx 1\text{--}2$. The total cache capacity can then be scaled to $N$ vectors by adding additional mini-indexes of size $w$. As a result, the cache lookup cost becomes nearly constant with respect to the cache size $N$ and is well-approximated by $c \log w$. %Moreover, as the working set changes, vectors from previous working sets become cold and are naturally evicted at the granularity of mini-indexes, which further reduces eviction-induced information loss.

%In summary, we recommend setting the total cache capacity $N$ (in vectors) large enough to accommodate the expected working set size, as is standard practice for any caching system, and generally larger. Due to the access skew described in Section~\ref{sec:workload-characteristics}, this is relatively inexpensive to provision. The working set size grows linearly with $k$ and with the number of diverse (semantically dissimilar) queries arriving within a caching interval (i.e., queries per window in Figure~\ref{fig:sliding-window}). A best-effort estimate of the working set can guide the choice of $c_{\text{mini-index}}$, as suggested in Section~\ref{sec:granularity-matters}. Even if this estimate is imperfect, recall and hit ratio remain largely unaffected (Figure~\ref{fig:cache_size_experiments}b), though hit latencies may increase.

\begin{figure}[h]
    \centering
    \begin{subfigure}{0.48\linewidth}
        \centering
        \includegraphics[width=\linewidth]{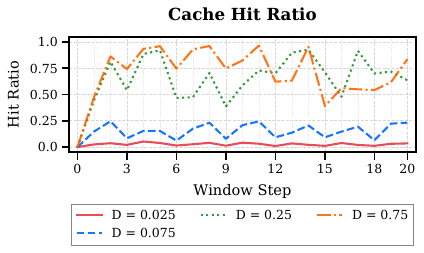}
    \end{subfigure}\hfill
    \begin{subfigure}{0.48\linewidth}
        \centering
        \includegraphics[width=\linewidth]{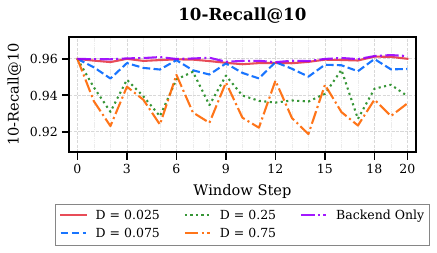}
    \end{subfigure}

    \caption{Deviation Factor Effect on Hit Ratio - Recall on SIFT.}
    \label{fig:deviation-factor-experiment}
    \vspace{-\baselineskip}
\end{figure}

\subsection{Controlling Recall and Cache Hit Ratio via Deviation Factor}
\label{sec:deviation-factor}
The hyperparameter $D$, the deviation factor, gives users explicit control over the trade-off between recall and hit ratio. To quantify its effect, we repeat the experiment from Section~\ref{sec:dataset-experiments} on SIFT while varying $D$. The results in Figure~\ref{fig:deviation-factor-experiment} show that increasing $D$ improves the cache hit ratio at the cost of reduced recall. In practice, this trade-off exhibits a saturation effect: beyond a certain point, further increases in $D$ yield only marginal gains in hit ratio while incurring only small additional recall loss.

We do not prescribe a single rule of thumb for choosing $D$. In our experiments in Section~\ref{sec:dataset-experiments}, we selected $D$ by starting from $0$ and incrementally increasing it by $0.025$ at each step while monitoring the resulting hit ratios. Once the hit ratio stopped improving meaningfully, we stopped increasing $D$. This procedure can be performed online at runtime without any downtime.

%As shown in Figure~\ref{fig:spatial-threshold-experiment}, the global threshold aggregates updates from cache misses across all regions, failing to capture local patterns and causing up to a 15\% loss in recall due to incorrect hit/miss decisions. In contrast, spatial thresholds adapt to local variations in the query distribution, limiting recall degradation to at most 2–3\%, demonstrating their effectiveness in preserving the backend database’s accuracy.

\begin{figure}[h]
    \centering

    % Row 1: Hit Ratio
    \begin{minipage}{0.48\linewidth}
        \centering
        \includegraphics[width=\linewidth]{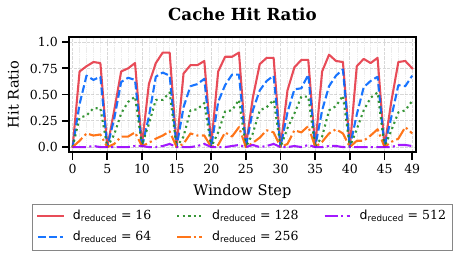}
        \subcaption{$d_{\text{reduced}}$ — Hit Ratio}
    \end{minipage}\hfill
    \begin{minipage}{0.48\linewidth}
        \centering
        \includegraphics[width=\linewidth]{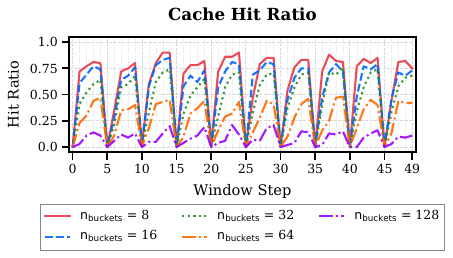}
        \subcaption{$n_{\text{buckets}}$ — Hit Ratio}
    \end{minipage}

    \vspace{0.8em}

    % Row 2: Recall
    \begin{minipage}{0.48\linewidth}
        \centering
        \includegraphics[width=\linewidth]{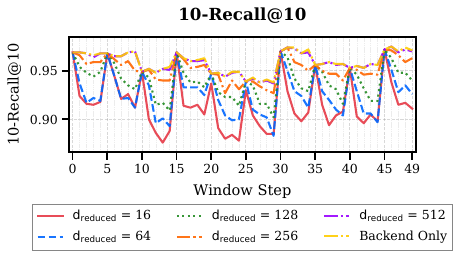}
        \subcaption{$d_{\text{reduced}}$ — Recall}
    \end{minipage}\hfill
    \begin{minipage}{0.48\linewidth}
        \centering
        \includegraphics[width=\linewidth]{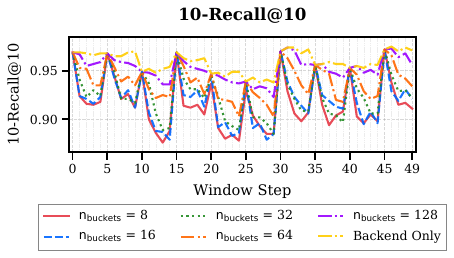}
        \subcaption{$n_{\text{buckets}}$ — Recall}
    \end{minipage}

    \caption{Impact of granularity of space partitioning and dimensionality reduction on recall and hit ratio on GIST. Left: varying $d_{\text{reduced}}$ (fixed $n_{\text{buckets}}$ to 8). Right: varying $n_{\text{buckets}}$ (fixed $d_{\text{reduced}}$ to 16).}

    \label{fig:pca_experiments}
    \vspace{-\baselineskip}
\end{figure}

\subsection{Sensitivity Analysis: Space Partitioning and Dimensionality Reduction}
\label{sec:space-partitioning}

We evaluate QVCache’s sensitivity to the granularity of space partitioning ($n_{\text{buckets}}$) and dimensionality reduction ($d_{\text{reduced}}$) by repeating the experiment from Figure \ref{fig:dataset-experiments} on the GIST dataset as it has the highest dimensional vectors. As shown in Figure \ref{fig:pca_experiments}, the left column fixes ($n_{\text{buckets}}$) at 8 and varies ($d_{\text{reduced}}$), while the right column fixes ($d_{\text{reduced}}$) at 16 and varies ($n_{\text{buckets}}$), reporting the resulting hit ratio and recall.

GIST vectors have 960 dimensions. Increasing ($d_{\text{reduced}}$) to 128 or higher provides only a modest improvement in recall (around 3–4\%) while significantly reducing the hit ratio, highlighting the effectiveness of dimensionality reduction for guiding cache hit–miss decisions.

Similarly, increasing ($n_{\text{buckets}}$) from 8 to 128 yields an average recall improvement of roughly 5\% but heavily reduces the hit ratio. This behavior arises because Algorithm \ref{alg:learn-threshold} overfits local patterns (i.e. the partitioning is so fine-grained that $\theta[k][R]$ learns almost a query-specific estimate of $\text{d}{_\text{backend}}[k]$ in each region) and fails to generalize across queries.

\begin{table}[h]
\centering
\captionsetup{skip=6pt}
\small % Reduces font size; use \footnotesize for even smaller
\renewcommand{\arraystretch}{1.0} % Reduced from 1.15
\setlength{\tabcolsep}{4pt}    % Reduced from 6pt

\begin{tabular}{|>{\centering\arraybackslash}m{2.2cm}|c|c|c|c|c|c|}
\hline
\diagbox[width=2.2cm,height=0.9cm]
{$c_{\text{mini-index}}$}{$n_{\text{mini-index}}$}
& 1 & 2 & 4 & 8 & 16 & 32 \\
\hline
3{,}125
& \cellcolor{gray!25}16
& \cellcolor{orange!25}24
& \cellcolor{yellow!25}37
& \cellcolor{green!25}62
& \cellcolor{blue!25}113
& \cellcolor{purple!25}219 \\
\hline
6{,}250
& \cellcolor{orange!25}18
& \cellcolor{yellow!25}34
& \cellcolor{green!25}56
& \cellcolor{blue!25}100
& \cellcolor{purple!25}189
& \\
\hline
12{,}500
& \cellcolor{yellow!25}23
& \cellcolor{green!25}58
& \cellcolor{blue!25}99
& \cellcolor{purple!25}183
& & \\
\hline
25{,}000
& \cellcolor{green!25}34
& \cellcolor{blue!25}108
& \cellcolor{purple!25}170
& & & \\
\hline
50{,}000
& \cellcolor{blue!25}56
& \cellcolor{purple!25}171
& & & & \\
\hline
100{,}000
& \cellcolor{purple!25}99
& & & & & \\
\hline
\end{tabular}

\caption{Memory usage (in MB) of QVCache with varying
$n_{\text{mini-index}}$ and $c_{\text{mini-index}}$ (in vectors) on SIFT.}
\label{tab:memory-footprint-experiment}
\vspace{-\baselineskip}
\end{table}

\subsection{Memory Overhead Analysis of QVCache}
\label{sec:memory-overhead}

For a billion-scale dataset such as SIFT, DiskANN requires 33.5GB of memory, and in-memory backends like FAISS can reach into the hundreds of gigabytes. By comparison, adding QVCache with a capacity of 100{,}000 vectors incurs only 100–200MB of additional memory. As expected, memory usage grows linearly with total cache capacity (and with vector dimensionality), and we observe that partitioning across multiple mini-indexes further increases overhead, as indicated by the diagonals in Table~\ref{tab:memory-footprint-experiment}. However, this overhead remains negligible relative to the memory consumed by the backends. Even with a very generous QVCache budget (e.g., 1M vectors), the additional cost for SIFT is only about 1–2GB.

The memory required to store distance thresholds is also negligible and is already included in the numbers reported in Table~\ref{tab:memory-footprint-experiment}. For example, in Figure~\ref{fig:dataset-experiments}, QVCache learns roughly 1.5K, 15K, and 50K thresholds for GIST, SIFT, and SpaceV, respectively, consuming about 200KB for SpaceV. %Even under pessimistic assumptions where, for instance, one million thresholds are learned due to highly diverse queries or varying values of $k$, the footprint remains around 4MB.%
Users may optionally cap the number of stored thresholds, evicting and relearning them if needed. %Overall, threshold storage contributes insignificantly to QVCache’s memory usage.

\begin{figure}[h]
    \centering

    % Top row
    \begin{subfigure}[t]{0.48\linewidth}
        \centering
        \includegraphics[width=\linewidth]{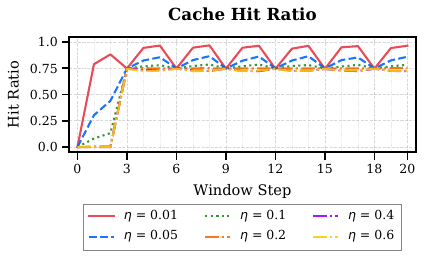}
    \end{subfigure}\hfill
    \begin{subfigure}[t]{0.48\linewidth}
        \centering
        \includegraphics[width=\linewidth]{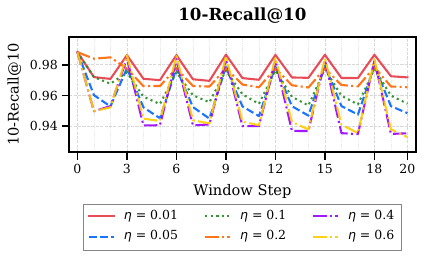}
    \end{subfigure}

    \vspace{0.75em}

    % Bottom row
    \begin{subfigure}[t]{0.48\linewidth}
        \centering
        \includegraphics[width=\linewidth]{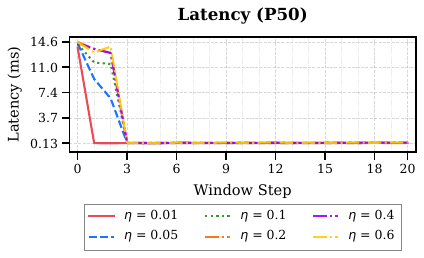}
    \end{subfigure}
    \begin{subfigure}[t]{0.48\linewidth}
        \centering
        \includegraphics[width=\linewidth]{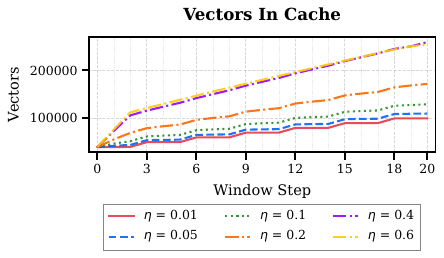}
    \end{subfigure}

    \caption{Effect of increasing query noise ratio, $\eta$, on DEEP dataset.}
    \label{fig:noise-ratio-experiments}
    \vspace{-\baselineskip}
\end{figure}

\subsection{Stress-Testing QVCache in the Absence of Temporal–Semantic Locality}
\label{sec:temporal-semantic-locality-experiments}

We next study the robustness of QVCache to different degrees of query perturbation~$\eta$. In Figure~\ref{fig:noise-ratio-experiments}, we evaluate the effect of~$\eta$ by repeating the experiment from Section~\ref{sec:dataset-experiments} while increasing $\eta$ up to~0.6. To avoid triggering evictions and focus solely on how many vectors QVCache retrieves into the cache, we set the cache capacity to $1$M vectors. As shown in Figure~\ref{fig:overlap-analysis}, $\eta = 0.6$ represents an extreme case in which perturbed queries share no overlap in their top-$k$ neighbors. Even under this setting, QVCache sustains high achieve ratios while degrading recall by less than~4\%.

This result reveals an important phenomenon: pairwise dissimilarity among queries does not imply global dissimilarity across a workload. Although no two perturbed queries share top-$k$ nearest neighbors, the collective set of vectors they reference still exhibits overlap at scale, i.e. under high query concurrency and volume. This is reflected in the curves for $\eta = 0.4$ and $\eta = 0.6$, where the number of vectors inserted into the cache increases only modestly. 
Concretely, although the workload issues 84{,}000 queries in total, the settings $\eta = 0.4$ and $\eta = 0.6$ lead to fewer than 300{,}000 vectors being inserted into the cache, far below the 840{,}000 vectors that would be expected in the absence of any pairwise top-$k$ neighbor overlap (with $k = 10$). Thus, QVCache may remain effective even when similar queries do not repeat, leveraging collaboration across many dissimilar queries (indicated by the cache hit ratio around 0.75 in Figure \ref{fig:noise-ratio-experiments}) rather than relying solely on temporal–semantic locality.

\section{Related Work}

\textbf{Vector Databases:} The growing demand for managing embedding data has led to the development of numerous vector database management systems in recent years \cite{pinecone, pgvector, ZillizServerless2025, milvus, qdrant2025, OpenSearch, 10.14778/3685800.3685806-gaussdb}. These systems incorporate a variety of optimizations tailored to vector data, including storage architectures, lock management, and query processing. %Moreover, several studies have explored accelerating ANN search using specialized hardware, including GPUs~\cite{bang,scalablegraphindexingusinggpus} and FPGAs~\cite{vector-search-delayed-fpga}. 
Furthermore, as vector data is increasingly combined with relational data, recent research has focused on supporting fundamental relational operations such as joins and filtering within vector databases, which are known as similarity joins \cite{Chen_2025} and filtered vector search \cite{10.14778/3750601.3750700}.

\textbf{Caching in Vector Databases:}  Caching in vector databases typically refers to system-level mechanisms \cite{jeong2025callcontextawarelowlatencyretrieval, milvus, 10.14778/3685800.3685805, turbocharging-vector-databases-ssds, diskann, tiered-cache-hnsw} that are tightly coupled with their respective systems and underlying indexes. These approaches primarily aim to reduce the cost of disk accesses arising from random I/O during graph traversal. For example, \cite{diskann} caches nodes near traversal entry points in memory, \cite{tiered-cache-hnsw} keeps the uppermost HNSW layer resident in memory, \cite{jeong2025callcontextawarelowlatencyretrieval} batches queries by aligning their page requests, and \cite{turbocharging-vector-databases-ssds} reorganizes the on-disk index layout to minimize page-cache misses.

\textbf{Similarity Caching:}
Similarity based caching has recently gained traction in the context of LLM APIs and document retrieval. The core idea is to place a query level cache in front of the model or retrieval engine: if an incoming prompt or query is sufficiently similar to a previously seen one, according to a similarity function and a predetermined threshold, the response is returned directly from the cache. Because conversational systems and RAG pipelines often exhibit substantial semantic repetition across queries \cite{10.14778/3750601.3750679, 10.1145/3578519, 10.14778/3685800.3685905, 10.1145/3721146.3721941, gptcache}, this strategy has proven highly effective. For example, \cite{gptcache} stores LLM prompts and responses to eliminate redundant API calls, while \cite{10.14778/3685800.3685905} caches retrieved document sets to accelerate subsequent retrieval queries. Additionally, a recent study \cite{vcache} proposed a method for semantic caching that provides formal guarantees on the error rate.

\section{Conclusion}
We introduced QVCache, a query-aware, backend-agnostic vector cache that achieves sub-millisecond cache-hit latencies independent of dataset size, while consuming only a memory footprint on the order of megabytes. By dynamically learning region-specific distance thresholds, QVCache delivers 40–1000$\times$ lower query latencies on cache hits compared to existing similarity search systems, without much compromising recall. Across diverse datasets and backend vector databases, QVCache consistently accelerates query execution by converting queries into cache hits, demonstrating that adaptive similarity caching is a practical and effective optimization layer for large-scale vector search systems, particularly under workloads exhibiting temporal-semantic locality.

\bibliographystyle{ACM-Reference-Format}
\bibliography{references}

\end{document}